\documentclass[twocolumn]{aa}

\usepackage[dvipsnames]{xcolor}
\usepackage{graphicx}
\usepackage[switch]{lineno}
\usepackage{verbatim}
\usepackage{pgfplots}
\usepackage{numprint}

\usepackage{txfonts}
\usepackage{listings}
\usepackage{graphicx}
\usepackage{textcomp}
\usepackage{upgreek}
\usepackage{longtable}
\usepackage[export]{adjustbox}
\usepackage{afterpage}
\usepackage{emptypage}
\usepackage[autostyle]{csquotes}
\usepackage{textcomp}
\usepackage{float}

\usepackage{enumitem}
\usepackage{subcaption}
\usepackage{tabu}
   \usepackage{booktabs}
   \usepackage{amsmath}
   \usepackage{amssymb}

   \usepackage{bm}
\usepackage{caption}
\usepackage[alsoload=abbr]{siunitx}
\usepackage{ulem}
\def\jerome#1{{\textcolor{blue}{\textbf{#1}}}}

\begin{document} 
\setpagewiselinenumbers
\nolinenumbers
   \title{The Galactic population of canonical pulsars}

   \author{Ludmilla Dirson 
          \inst{1},
          Jérôme Pétri, \inst{1}
         and Dipanjan Mitra \inst{2,3}
          }

    \institute{Universit\'e de Strasbourg, CNRS, Observatoire astronomique de Strasbourg, UMR 7550, F-67000 Strasbourg, France.\\
              \email{dirson@astro.unistra.fr}   
 \and 
National Centre for Radio Astrophysics, Tata Institute for Fundamental Research, Post Bag 3, Ganeshkhind, Pune 411007, India
\and
Janusz Gil Institute of Astronomy, University of Zielona G\'ora, ul. Szafrana 2, 65-516 Zielona G\'ora, Poland     
}

   \date{Received ; accepted }


  \abstract
   {Current wisdom accounts to the diversity of neutron star observational manifestations to their birth scenarios, influencing their thermal and magnetic field evolution. Among the kind of observed neutron stars, radio pulsars represent by far the largest population of neutron stars.}
   {In this paper, we aim at constraining the observed population of the canonical neutron star period, magnetic field and spatial distribution at birth in order to understand the radio and high-energy emission processes in a pulsar magnetosphere. For this purpose we design a population synthesis method self-consistently taking into account the secular evolution of a force-free magnetosphere and the magnetic field decay.}
   {We generate a population of pulsars and evolve them from their birth to the present time, working in the force-free approximation. We assume a given initial distribution for the spin period, surface magnetic field and spatial galactic location. Radio emission properties are accounted by the polar cap geometry, whereas the gamma-ray emission is assumed to be produced within the striped wind model.}
   {We found that a decaying magnetic field gave better agreement with observations compared to a constant magnetic field model. Starting from an initial mean magnetic field strength of $B=2.5\times 10^8$~T with a characteristic decay timescale of $4.6 \times 10^5~$yr, a neutron star birth rate of $1/70~$yr and a mean initial spin period of 60~ms, we found that the force-free model satisfactorily reproduces the distribution of pulsars in the $P-\dot{P}$ diagram with simulated populations of radio-loud, radio-only and radio quiet gamma-ray pulsars similar to the observed populations.}
   {}


    \maketitle
\textbf{}%



\section{Introduction}


Pulsars are rapidly rotating, highly magnetized neutron stars surrounded by a plasma filled magnetosphere that emits regular pulses of radiation at their spin frequency. The term \textit{pulsar} comes from 'pulsating radio source' since they were first observed at radio wavelengths. Currently, the number of observed radio pulsars is roughly~3,000 \footnote{according to the ATNF catalogue at \url{https://www.atnf.csiro.au/research/pulsar/psrcat/}} \citep{atnfCat}. Subsequently, however it was found that pulsars were also bright in X-rays, optical and gamma-rays. In particular, the Large Area Telescope (LAT) on board of the Fermi satellite has forcefully changed our understanding of gamma-ray pulsars by discovering dozens of radio-quiet gamma-ray pulsars as well as millisecond pulsars (MSPs). After more than 10~years of operation, the LAT has detected over 250~pulsars (\cite{Fermi_2FGL}; \cite{Smith2019}), establishing those as the dominant class of GeV sources in the Milky Way. This putted the gamma-ray pulsar statistics to a new era, first opened by EGRET with only 7~pulsars \citep{thompson_second_1995}. These recent discoveries have raised fundamental questions, one of the most challenging amongst them is to find a physical consistent theory of pulsar multi-wavelength emission mechanisms. With the advancement of instrumental techniques, future surveys with greatly improved sensitivities are being envisaged. Amongst them, the Square Kilometre Array (SKA) is expected to be 50~times more sensitive, and is predicted to survey the sky 10,000~times faster than any existing imaging radio telescope array \footnote{\url{https://www.skatelescope.org/wp-content/uploads/2011/03/SKA-Brochure_web.pdf}}. In this context, it is of utmost importance to be able to anticipate the discovery rate of new pulsars, since it enables to validate our current understanding of pulsar emission mechanisms, as well as their evolution track and the factors impacting their detection. 


Pulsar population synthesis (hereafter PPS) studies are powerful tools aiming to predict the detectability of pulsars. Inferring the underlying properties responsible for the observed pulsar population is an arduous task. Indeed, both the impact of the detection procedure and the inhomogeneous properties of the interstellar medium have to be accurately estimated to account for possible observational biases. 

Much of the advance in pulsar population synthesis has come from thorough Monte Carlo simulations that generate pulsars and test whether they fulfil the criteria for detection according to geometrical factors and sensitivity issues. It is then possible to develop and optimize a model for the underlying pulsar population, which informs us about the important intrinsic neutron star parameters and distribution, enabling predictions for future surveys. Usually, PPS studies follow two simple approaches. The first one is to take a 'snapshot' of the Galaxy as it appears today, where no assumptions are made regarding the prior evolution of the pulsar population. Instead, this population is generated assuming various distribution functions (typically spatial distribution, spin period~$P$ and  $\dot{E}$- or $\dot{P}$) which are optimized to match the observations. Inspired by earlier studies from \cite{Taylor_Manchester1977}, \cite{Lyne_Manchester1985} and \cite{Parkes_multibeam_pulsar_survey2006} have applied the snapshot approach to the canonical \footnote{Defined here as the pulsars that are non-binaries, with $P>20$~ms and that are not magnetars.} pulsar population to determine best-fitting probability density functions in Galactocentric radius ($R$), luminosity ($L$), height with respect to the galactic plane ($z$) and the period~$P$ for the currently observed population of pulsars. Alternatively, one may consider 'evolution' strategies where the pulsars are evolved from birth up to the present era starting from an initial spatial distribution and an initial period and magnetic field distribution. A fine example of the latter genre is the comprehensive study of \cite{FaucherGiguere_Kaspi2006}, which quite successfully reproduced the properties of the main part of the radio pulsar population using a model in which the luminosity has a power-law dependence on $P$ and $\dot{P}$. PPS studies can be also extended to the population of neutron stars observed in other bands such as X-rays (see for instance \cite{Popov}) and $\gamma-$rays.
Indeed, with the broad increase in $\gamma$-ray pulsar numbers, a statistical treatment of the $\gamma$-ray population in combination with deep radio surveys of the galactic plane is now feasible. Early works on radio-loud gamma-ray pulsar populations carried out before the Fermi era include \cite{Gonthier2002}, 2004, 2007a, 2007b. With the advent of Fermi/LAT new studies emerged (see for instance \cite{Gonthier2018}; \cite{Ravi2010}; \cite{Takata_OGmodel1}; \cite{WattersRomani}; \cite{Pierbattisata2010}), trying to constrain the geometry and the location of the gamma-ray emission sites. \cite{WattersRomani} showed that an initial spin period of $P_0=50~$ms and a birth rate of 1 neutron star per 59~yr were required to reproduce the observed $\gamma$-ray population. They made the prediction that after 10~years of operations Fermi should detect $\sim$120~young $\gamma$-ray pulsars of which about one half would be radio-quiet. \cite{Gonthier_beam_pulse_profiles2004} included an exponentially decaying magnetic field with a 2.8~Myr timescale and displaying a satisfactory agreement with the $P-\dot{\mathrm{P}}$ distribution at that time. Later, more accurate magnetic field decay models have been elaborated, especially for magnetars as presented in \cite{vigano_2013}.

The gamma-ray emission modelling also drastically benefited from the second Fermi pulsar catalogue. Based on fluid simulations in dissipative regime by extension of the force-free regime, \cite{kalapotharakos_gamma-ray_2014} constrained the magnetic axis inclination with respect to the rotation axis and computed the associated light-curves for curvature radiation. Later, \cite{Kalapotharakos2017} constrained also the dissipation rate depending on the pulsar spin down, differentiating millisecond pulsars from young pulsars. Eventually, \cite{Kalapotharakos2018_3D} gave a simple analytical fit for the gamma-ray luminosity depending on the cut-off energy, spin down and surface magnetic field. We will use these results to predict the gamma-ray flux in our population synthesis.

The appropriate emission mechanisms of $\gamma$-rays emission from pulsars are still under active investigation. Various models have been proposed to explain the particle acceleration sites as the origin of pulsed $\gamma$-rays emission. The first one suggested is the polar cap region, which is confined in the open magnetosphere at low altitudes (\cite{sturrock_model_1971}; \cite{Ruderman_Sutherland1975}; \cite{Daugherty_Harding1982}; \cite{Daugherty_Harding1996}). The slot-gap (along the last open magnetic field lines, \cite{arons_pair_1983}; \cite{Muslimov_2004}; \cite{Harding_slot_gap2008}; \cite{Muslimov_Harding2011}) and the outer-gap model (extending to the edge of the light cylinder, \cite{Cheng_Ruderman_a}, 1986b; \cite{Hirotani_2008}, \cite{Takata_OGmodel1}) both located at high altitudes in the outer magnetosphere. Polar cap models predict a sharp, super-exponential cut-off at several GeV, which is steeper than those of the $\gamma$ -ray spectra of the Crab and Vela pulsars measured by Fermi (\cite{Fermi_2FGL}). At the beginning of the years~2000 a new picture emerged for the production site of pulsars gamma-rays. It is located well outside the light-cylinder in the striped wind \citep{kirk_pulsed_2002} based on the structure discussed by \cite{coroniti_magnetically_1990} and by \cite{michel_magnetic_1994}. Relativistic beaming and the spiral structure of the emitting current sheet produces a pulsed radiation (\cite{Petri2009}, \cite{Petri2011}. No PPS studies have been carried out so far by considering the latter emission site. Furthermore, although pulsar magnetospheres are filled with a relativistic electron-positron plasma (\cite{Goldreich_Julian1969}), most of the work on pulsar population studies are carried out assuming a spherical neutron star rotating in vacuum. However, the radiation from neutron stars is produced by charged particles flowing within their magnetosphere at ultra-relativistic speeds. It is therefore necessary to take the plasma back reaction into account for the pulsar period and magnetic inclination angle evolution. State-of-the-art simulations of the magneto-thermal evolution \citep{vigano_2013} and revised magnetospheric models \citep{Philippov} allow for a more accurate prediction of the long-term behaviour of magnetic field strength, angular momentum loss rate and the inclination angle (angle between the magnetic dipole moment and the rotational axis). Following these studies, \cite{Gullon2014} have revisited the population synthesis of isolated radio-pulsars incorporating a magnetic field decay in the framework of the vacuum approximation and realistic magnetospheric models to be compared with each other. 
Following the work of \cite{Gullon2014}, we develop an evolution model for the population study of canonical pulsars in both radio and gamma-ray bandwidth by taking into account both the evolution of the dipole magnetic inclination angle and the decay of the magnetic field in the force-free approximation. The striped wind model is used for the first time in a PPS study to describe the observed $\gamma$-ray pulsar population. We use a polar cap model for radio emission. We compare the results of our simulations with the observations by using the shape of the $P-\dot{P}$ diagram, as well as the number of radio-only, gamma-only and gamma-ray radio loud pulsars (see Table \ref{tab:obs_Ndetec}).
In order to get better constrain on the radio emission sites from the polarization data following the rotating vector model, we focus only on young (non-recycled) pulsars for which the radio emission height is reasonably well constrained.


\begin {table}[H] 
\centering
\begin{tabular}{l l l l l l}
 \hline
 \hline
  $\log(\dot{E})$ (in W) &$N_{\rm tot}$ & $N_{\rm r}$ & $N_{\rm g}$ & $N_{\rm rg}$ \\  
     \hline
  
$> 31 $ & 2 &  0 & 0  & 2   \\
$> 28 $ & 197&  101 & 35 & 61  \\
 \hline 
 \hline
 total & 2665  & 2553  & 63  & 84 \\
\hline
\end{tabular}
\caption[caption for LOF]{The number of known pulsars with $\dot{E}$ above and below $10^{28}$~W and above $10^{31}$~W. \textsuperscript{4} }
 \small\textsuperscript{4} The quantities $N_{\rm r}$, $N_{\rm g}$, $N_{\rm rg}$ are the number of radio only, gamma only and radio-loud gamma-ray pulsars respectively. Note that we have excluded the binary pulsars as well as pulsars with $P<20~$ms. The data have been taken from the ATNF catalogue and from \protect\url{https://confluence.slac.stanford.edu/display/GLAMCOG/Public+List+of+LAT-Detected+Gamma-Ray+Pulsars}
\label{tab:obs_Ndetec}
\end {table}  





The paper is organised as follows. In Section~\ref{sec:model} we outline the model that we used to generate the observed pulsar population, whereas their detection is addressed in section~\ref{sec:detection}. In section~\ref{sec:simulations}, we present the results of our simulations and discuss their signification in section~\ref{sec:discussion}. A summary is proposed in section~\ref{sec:summary}.


\section{$P-\dot{P}$ evolution model }
\label{sec:model}

We start our population synthesis analysis by describing the underlying model of the period and luminosity evolution starting from an initial sample of neutron stars with magnetic field strength~$B_0$ and birth period~$P_0$. 
We use a Gaussian distribution for the initial spin-period and a Gaussian in log space distribution for the magnetic field, such that
\begin{equation}
p(\log(B_0)) =\frac{1}{\sigma_b \sqrt{2\pi}}e^{-(\log B_0 - \log \bar{B})^2/(2\sigma_b^2)} 
\end{equation}
We assume mean values of $\Bar{P}= 60$~ms and $\Bar{B}= 2.5 \times 10^{8}$~T, and standard deviation $\sigma_{p}=10$~ms and $\sigma_{b}=0.5$. 
These distributions are similar to the ones used by \cite{Gullon2014}, \cite{D.Smith}, \cite{FaucherGiguere_Kaspi2006}, \cite{yadigaroglu_gamma-ray_1995} and  \cite{WattersRomani}. We also assume an isotropic distribution of magnetic inclination angles~$\alpha$ with respect to the rotation axis, meaning that the variable $\cos \alpha$ is uniformly distributed between 0 and 1.


Moreover, in our model, we generate a population of pulsars with a constant birth rate of $1/70~\mathrm{yr}$. Note the age of each pulsar should be taken uniformly from age zero up to the age of the Milky Way that is about the age of the Universe \numprint{13e9}~yr, hence a total number of \numprint{1.9e8} pulsars should normally be simulated. Here we choose to generate only $10^7$ pulsars since a higher number of simulated pulsars such as $10^8$ pulsars increase the simulation time significantly. However, we have checked that in a few instances of simulations using $10^8$ pulsars only changes the final results only by a few percent.
We include in our model a magnetic field decay with a decay rate $\tau_d = k_{\tau_d} \cdot \tau_{v}$ with $\tau_v$ the decay rate given by \cite{vigano_2013} by extrapolation to lower field strength compared to the magnetars they studied. The pulsar, then evolves with a decreasing total spin down power~$\dot{E}$ which serves as a proxy for the radio and gamma-ray luminosity as shown later.

In order to derive the pulsar's detectability, we need to know its distance~$d$ to the Earth. To summarize, each pulsar has the following intrinsic characteristics
\begin{itemize}
    \item $B_{0}$, magnetic field at birth.
    \item $P_{0}$, period at birth.
    \item $P$ and $\dot{P}$, current period and its time derivative.
    \item $\alpha_{0}$, inclination angle at birth taken isotropically  (uniform in $\cos \alpha_{0}$).
    \item $\alpha$, current inclination angle.
    \item $\vec{n_{\Omega}}$, unit vector along the rotation axis.
    \item $(x_0, y_0, z_0)$ birth position in Galactocentric coordinates.
    \item $v_{0}$ birth kick velocity.
    \item $d$, distance to Earth.
\end{itemize}\textbf{}
Pulsars are spread all over the Galaxy with a radial and height distribution above the galactic plane deduced from observations. In a Cartesian coordinate system attached to the galaxy, they are located at individual positions denoted by $(x,y,z)$. We will try to retrieve the pulsar current spatial distribution from an initial distribution where they are concentrated in the galactic plane. 


\subsection{Birth and evolution of the pulsars}

The current period~$P$ and magnetic moment inclination angle~$\alpha$ are evolved from their initial values at birth, according to some spin evolution models.
We consider a force-free evolution scenario, as well as the evolution of the inclination angle~$\alpha$. The parameters that we used in our simulations are summarized in table~\ref{tab:params}.
\begin {table}[H] 
\small\tabcolsep=0.15cm
\begin{tabular}{ l l l l l l l l}
 \hline
 \hline
$\tau_{birth}$ (1/yr) & $P_{mean}~$(ms) & $B_{mean} $ (T) & $\sigma_{P}~$(ms) & $\sigma_{B}$ & $\alpha_d$ & $k_{\tau_d}$ \\  
       \hline
70 & 60  & $2.5 \times 10^8$   & 10 & 0.5  & 1.5 &  5 & \\
\hline
\end{tabular}
\caption{Parameters used in our simulations. }
\label{tab:params}
\end {table}  

\subsubsection{Vacuum dipole}

The most studied rotator is a magnetic dipole radiating in vacuum. Exact solutions have been given by \cite{deutsch_electromagnetic_1955} but the point dipole formula is sufficient to accurately evolve the period \citep{jackson_electrodynamique_2001}. The spin down luminosity is given for the orthogonal rotator by
\begin{equation}
\label{eq:Lperp}
L_{\perp}=\dfrac{8 \pi}{3 \mu_{0}c^3}B^{2} \Omega^{4} R^{6}
\end{equation}
and for an oblique rotator by
\begin{equation}
\label{eq:Ldip}
L_{\rm dip} = L_\perp \, \sin^2 \alpha
\end{equation}
where $\alpha$ is the magnetic obliquity, $B$ the magnetic field strength at the magnetic equator, $\Omega=2\pi/P$ the rotation speed, $R$ the neutron star radius and $\mu_{0}$ the vacuum permeability which has the value $\mu_{0}=4\pi \times 10^{-7}~$H/m.

This spin down removes energy and angular momentum from the star, leading to a braking with an increase of the period at a rate $\dot{P}$ 
related to the rate of rotational kinetic energy loss~$\dot{E}$ such that
\begin{equation}\label{Edot}
\dot{E} = \frac{{\rm d}E_{rot}}{{\rm d}t} = - I \Omega \dot{\Omega} = L_{\rm dip}
\end{equation}
where $\dot{\Omega}$ is the spin frequency derivative. The canonical value of the moment of inertia is $I = 10^{38}~\mathrm{kg~m^2}$.
Combining equation~\eqref{eq:Ldip} and eq.~\eqref{Edot}, the expected evolution of the rotation frequency becomes 
\begin{equation}\label{dotOmega}
\dot{\Omega}= \dfrac{8 \pi}{3\mu_{0}c^3}\dfrac{(BR^3 \sin \alpha)^{2}}{I} \Omega^3 .
\end{equation}
The spin-down of a pulsar is generally written in a more concise form
\begin{equation}
\dot{\Omega} =-K_{\rm vac} \, {\Omega}^{n},
\end{equation}
where $n$ is the braking index. From the equation~\eqref{dotOmega}, we deduce its value to be $n=3$ which also holds for a force-free magnetosphere, see for instance \cite{petri_general-relativistic_2016}, and 
\begin{equation}
 K_{\rm vac} = \dfrac{8 \pi}{3\mu_{0}c^3}\dfrac{(BR^3 \sin \alpha)^{2}}{I}=K_1 B^2 \sin^2 \alpha .
 \end{equation}
with $(K_1=\d8 \pi R^6)/(3\mu_{0}c^3 I)$
For a typical neutron star radius of $R=12~$km as found by recent NICER observations \citep{riley_nicer_2019, bogdanov_constraining_2019} this constant is estimated to
\begin{equation}
K_{\rm vac} \approx \SI{e-15}{\second^{-1}} \, \left( \frac{B}{\SI{4.5e8}{\tesla}}\right)^2 \, \left( \frac{R}{\SI{12}{\kilo\meter}}\right)^6 \, \left( \frac{I}{\SI{e38}{\kilogram\square\meter}}\right)^{-1} \, \sin^2 \alpha .
\end{equation}

From Equation ~\eqref{dotOmega}, assuming a constant obliquity~$\alpha =  \alpha_0$ we can derive the period~$P$ evolution as
\begin{equation}\label{period}
P(t) =2 \pi \sqrt{2 \left(K_{\rm vac} \, t+\dfrac{P_{0}^{2}}{8 \pi^2}\right)} 
\end{equation}
From Eq.~\eqref{dotOmega}, we derive $P \, \dot{P}=4 \pi^2 \, K_{\rm vac}$ and $\dot{E}=4 \pi^2 I \dot{P}P^{-3}$ from Eq.~\eqref{Edot}.
These last equations, plot with straight lines in the $P-\dot{P}$ diagram and are usually shown in the approximation of vacuum field and no inclination angle evolution. The real path of a single pulsar can significantly deviate from the line of constant $B$.

\subsection{Force-free}

The vacuum model can be adapted to the force-free model by replacing the vacuum spin down by its force-free counterpart, as given by \cite{Spitkovsky2006} and \cite{petri_pulsar_2012}
\begin{equation}
\label{Ldip}
L_{\rm ffe} = \frac{3}{2} \, L_\perp \,  (1+\sin^2\alpha) .
\end{equation}
The constant $K$ for the spin evolution becomes then
\begin{equation}\label{Eq:kffe}
K_{\rm ffe}= \frac{3}{2}K_{1} B^2(1+\sin^2\alpha) .
\end{equation}
The period then also follows \eqref{period} but with $K_{\rm vac}$ replaced by $K_{\rm ffe}$.

\subsection{Evolution of the inclination angle}

Concomitantly, the magnetic axis tends to align with the rotation axis with approximately the same timescale as the spin down. The stellar braking and spin alignment must therefore be evolved consistently in accordance with the electromagnetic torque exerted on its surface. 

A simple analytic solution for the time evolution of $\alpha$ for a vacuum rotator was found by \cite{michel_goldwire1970} who determined an exponentially fast alignment according to 
\begin{equation}\label{Eq:alpha_vacuum}
 \sin \alpha(t) = \sin \alpha_{0} \exp \left( -t/\tau^{\rm vac}_{\rm align}\right),
\end{equation}
  where $\tau^{\rm vac}_{\rm align}=1.5 \tau_{0} \cos^{-2} \alpha_{0}$ is the alignment timescale of a vacuum pulsar and
\begin{equation}
\tau_{0}=\frac{Ic^3}{\mu^2 \Omega_{0}^2} \approx 10^4 \left( \frac{B_{0}}{10^{8}~ \mathrm{T}}\right)^{-2} \left( \frac{P_{0}}{10~ \mathrm{ms}} \right)~\mathrm{yr^{-1}}
\end{equation}
is the characteristic spin-down time of a pulsar, with $\mu$ the norm of the magnetic moment. The subscript '0' refers to the values of pulsar variables at birth, $t = t_{0}$ . 
Thus, the vacuum pulsar evolves to the aligned configuration exponentially fast, without a significant slow-down of its rotation.
When almost alignment is reached, due to the integral of motion $\Omega \cos \alpha = \Omega_0 \cos \alpha_0$, the asymptotic rotation rate becomes $\Omega = \Omega_0 \cos \alpha_0$ which is not significantly different from $\Omega_0$ for high initial inclination.

For a force-free magnetosphere, the situation is radically different. \cite{Philippov} indeed investigated the evolution of non-spherical pulsars with a plasma filled magnetosphere with the help of magneto-hydrodynamic (MHD) simulations. 
An analytic solution to the evolution of pulsar obliquity~$\alpha(t)$ derived by \cite{Philippov} is
\begin{equation}\label{Eq:f_alpha}
\frac{1}{2 \sin^2 \alpha(t)}+\log(\sin \alpha(t))=\frac{t}{\tau^{\rm MHD}_{\rm align}}+\frac{1}{2 \sin^2 \alpha_{0}}+\log(\sin \alpha_{0})
\end{equation}  
where $\tau^{\rm MHD}_{\rm align}=\tau_0 \sin^2 \alpha_0/\cos^4\alpha_0$ is MHD pulsar alignment timescale. The inclination angle asymptotes at later times to a power law, $\alpha(t) \propto t^{-1/2}$. The alignment timescale is therefore much longer than for a vacuum model. The actual $\dot{P}$ values are drastically different. Whereas in the vacuum case the spin down vanishes for an aligned rotator, for an MHD model, even the aligned rotator spins down. Consequently the period derivative~$\dot{P}$ decreases much slower for the latter model. 
The integral of motion is now
\begin{equation}
    \Omega \frac{\cos^2 \alpha}{\sin \alpha}=\Omega_0 \frac{\cos^2 \alpha_0}{\sin \alpha_0}
 \end{equation}
and contrary to the vacuum case, the final asymptotic rotation rate always tends to zero.

\subsection{Magnetic field decay}

The neutron star magnetic field is known to decay with time, depending on the temperature and strength of the field. In order to take this effect into account in our population synthesis, for simplicity, let us assume that the magnetic field decays according to a power law
\begin{equation}
\frac{{\rm d}B}{{\rm d}t} = - a \, B^{1+\alpha_d}
\end{equation}
where $a$ and $\alpha_d$ are constant parameters controlling the speed of the magnetic field decay. They are assumed to be independent of the neutron star model. Integrating in time, the magnetic field evolves as
\begin{equation}
B(t)=B_0(1+t/\tau_d)^{-1/\alpha_d}
\end{equation}  
with the initial magnetic field strength $B_0$ and the decaying timescale $\tau_d = 1/(\alpha_d \, a \, B_0^{\alpha_d})$. We notice that since $a\,\alpha_d$ is a constant, this decaying time depends on $B_0$. Thus, for a different initial magnetic field strength $B_1$ this timescale changes to
\begin{equation}\label{tau_B}
\tau_1 \, B_1^{\alpha_d} = \tau_d \, B_0^{\alpha_d} . 
\end{equation}

For a decaying field, equation~\eqref{Eq:f_alpha} is no more valid. It must be replaced by
\begin{multline}
\ln(\sin \alpha_0) +\frac{1}{2 \sin^2\alpha_0} + K \Omega_0^2 \frac{\cos^4 \alpha_0}{\sin^2 \alpha_0} \frac{\alpha_d \, \tau_d B_0^2}{\alpha_d-2} \left[ \left( 1+ \frac{t}{\tau_d} \right)^{1-2/\alpha_d} -1  \right]\\
    = \ln(\sin \alpha) +\frac{1}{2 \sin^2\alpha}
\end{multline}  
If the decaying is very slow, $\tau_d \gg t$ and the equation simplifies into \eqref{Eq:f_alpha}. 
The magnetic field decay timescale is affected by the initial field strength and the current age of each pulsar, as given by eq.~\eqref{tau_B}. The typical decay timescale for a mean magnetic field of $2.5 \times 10^8~$T is $4.6 \times 10^5~$yr which is consistent with the estimate of  \cite{vigano_magnetic_2013} for the same field strength. However, pulsar with magnetic field lower than the mean value will have longer decaying timescales and vice versa. Moreover, since we use equation~\eqref{tau_B} a value for $a$ is not needed. 

\subsection{Galactic distribution}

It is usually agreed upon that the z-distribution above the galactic plane of the population~I object is approximately exponential \citep{binney_merrifield}. The scale height of the z-distribution of pulsar progenitors is of the order $50-100$~pc (\cite{Mdzinarishvili}). Today, this disparity is explained by the fact that pulsars are moving away from their initial birthplace due to a large kick velocity.

For describing the position of neutron stars within the galaxy we work in the right-handed Galactocentric coordinate system $(x,y,z)$ that has the Galactic centre at its origin, with $y$ increasing in the disk plane towards the location of the Sun, and $z$ increasing towards the direction of the North Galactic Pole. In order to calculate the distance $d$ between the Earth and the pulsar, we need to know the Sun’s position with respect to the Galactic centre and with respect to the Galactic plane. 
As of today, neither of these quantities are absolutely known. The Sun is thought to be located at about $z_{\odot}=15~$pc (\citep{Siegert_sun_position}) above the Galactic plane and about $y_{\odot}=8.5~$kpc from the Galactic centre.
The distance of a pulsar from us is therefore $d = \sqrt{(x-x_\odot)^2+(y-y_\odot)^2+(z-z_\odot)^2}$. We will also use the cylindrical coordinate system ($R,\theta,z$), with $R$ the distance to an axis passing through the galactic centre in the plane of constant $z$. The Cartesian coordinates~$(x,y,z)$ are related to the cylindrical coordinates~$(R, \theta, z)$ in the usual way
\begin{equation}
  \begin{cases}
x = R \cos \theta \\
y =  R \sin \theta .
\end{cases}
\end{equation}

\subsubsection{Initial distribution}

The initial location of the pulsars is given by the distributions found by \cite{Paczynski_distributions_1990} for the radial direction
\begin{equation}  
\rho_{R}(R)=\frac{a_{R}e^{-R/R_{exp}}R}{R^2_{exp}}    
\end{equation}
and for the altitude by 
\begin{equation}
\rho_{z}(z)=\frac{1}{2z_{exp}}e^{-|z|/z_{exp}}    
\end{equation}
where $R$ is the axial distance from the $z-$axis, and $z$ is the distance from the Galactic disc. Numerical values are taken to be, $R_{exp}=4.5~$kpc and $a_{R}\equiv [  1-e^{1-R_{max}/R_{exp}}(1+R_{max}/R_{exp})]^{-1}=1.0683$ with $R_{max}=20~$kpc. Following \cite{vanderKruit1987}, we take $z_{exp}=75~$pc. 


\subsubsection{Birth kick velocity}

To describe the supernova kick velocity, we proceed similarly to \cite{Hobbs2005}, and use a Maxwellian distribution with a characteristic width of $\sigma_{v_{young}}=265~\mathrm{km~s^{-1}}$ for pulsars with age $< 3~$Myr and $\sigma_{v_{old}}=75~\mathrm{km~s^{-1}}$ for older ones. Therefore, One can write the element of velocity space as ${\rm d}^3 v={\rm d}v_{x}  {\rm d} v_{y} {\rm d} v_{z}$, for velocities in a standard Cartesian coordinate system. In this case, the distribution for a single direction is a normal distribution with mean $\mu_{v_{z}}=\mu_{v_{x}}=\mu_{v_{y}}=0$ and standard deviation $\sigma_{v_{x}}=\sigma_{v_{y}}=\sigma_{v_{x}}$. The mean velocity in three dimensions and the velocity dispersion in one dimension are connected by $\bar{v} \sim  \sigma_{v} \sqrt{8/\pi} \simeq 1.6 \sigma_{v}$.


\section{Detection}
\label{sec:detection}

In order to see if our generated set of pulsars is responsible for the observed set, we need to determine if they are actually detected. The pulsar detection is governed by three main factors.  
The first is the \textit{beaming fraction} which indicates the fraction of the sky covered by the radiation beam (in a particular wavelength, here we suppose radio or gamma-rays) and thus the fraction of the population that is possibly detectable from Earth. The beaming fractions of radio and $\gamma$-ray pulsars differ from each other and depend on the pulsar’s spin, geometry and location of the emission regions. In this paper, we use the force-free model to calculate the $\gamma-$ray beaming fraction. The second and third factors are the pulsar's \textit{luminosity} and the \textit{sensitivity} of a given radio or $\gamma-$ray survey. Indeed, the detection of pulsars depends on their brightness, location and on the sensitivity of the instruments. This requires an emission model for a pulsar. The most basic requirement of such a model is the presence of a dense plasma around the strongly magnetized neutron star. The plasma streams along the open magnetic field line regions and the radio emission rises close to the neutron star surface, while the gamma ray emission is constrained to originate close to the light cylinder and preferably outside it, within the striped wind. When combined, these two factors indicate if a given pulsar can be actually detected by a given instrument based on the geometry. To calculate the luminosity of each pulsar, we need its distance $d$ from us.
These three factors all together provide us with the actual number of detected pulsars.  

\subsection{Beaming fraction}

We assume isotropic distributions for the Earth viewing angle~$\xi$ and a conventional width of the radio beam denoted by $\rho$. The orientation of the unitary rotation vector ~$\vec{n_{\Omega}} = \dfrac{\vec{\Omega}}{\Vert \vec{\Omega} \Vert}$ is assumed to be isotropic, meaning uniform in $\phi$ and in $\cos \theta$. The cartesian coordinates of the unit rotation vector $\vec{n_{\Omega}}$ are $\left( \sin \theta_{n_{\Omega}} \cos \phi_{n_{\Omega}},  \sin\theta_{n_{\Omega}} \sin \phi_{n_{\Omega}},\cos  \theta_{n_{\Omega}} \right)$. 
 $\vec{n}$ is the unit vector along the line of sight, with coordinates:
\begin{equation}
\left( \dfrac{x-x_\odot}{d}, \dfrac{y-y_\odot}{d}, \dfrac{z-z_\odot}{d} \right) .
\end{equation}
When $\vec{n}$, $\bold{\mu}$ and $\vec{n_{\Omega}}$ are in the same plane, we can define the line of sight inclination angle as $\xi=( \widehat{\vec{n_{\Omega}},\vec{n}} )$, the magnetic axis inclination angle as $\alpha=( \widehat{\vec{n_{\Omega}},\vec{\mu}} )$. From these angles, we deduce the usual minimum angle between magnetic axis and line of sight as $\beta=( \widehat{\vec{\mu},\vec{n}} )$. 
The \textit{impact angle}, $\beta$ represents the closest path of the line of sight to the magnetic axis.

\subsection{Radio emission model}

\subsubsection{Radio beam geometry}

The radio beam geometry is explained in depth, for instance, in  Chapter~3 of \cite{lorimer_handbook_2004}. We summarize useful results in this section.
A simple geometrical relationship between the half opening angle of the emission cone ($\rho$), the emission height ($h_{em}$) and the spin period is given via
\begin{equation} \label{eq:rho}
\rho = 3~ \sqrt{\dfrac{\pi~ h_{em}}{2~P~ c}}~~\mathrm{radians}
\end{equation}
Equation ~\eqref{eq:rho} is well verified for slow pulsars \citep{lorimer_handbook_2004}.

The emission height $h_{em}$ is roughly constant, with a mean value of $h_{\rm em} \sim 3.10^5$~m across the population. This can be inferred from observations via Eqn.~\eqref{eq:rho}, as a number of studies found consistently $\rho  \propto P^{-0.5}$ \citep{mitra_nature_2017}.
With knowledge of the height $h_{em}$, the period distribution $P$ and $\alpha$ for each pulsar, the radio beaming fraction can then be computed for all simulated pulsars. 

The pulsar is detectable in radio if $\beta = |\xi - \alpha| \leq \rho$ (corresponding to the cone located in one hemisphere) or if $ |\xi -(\pi-\alpha)| \leq \rho $ if it is located in the other hemisphere. We must also satisfy $\alpha \geq \rho$ and $\alpha \leq \pi-\rho$ in order to effectively see radio pulsation. Because the line of sight must go in and out from the emission cone to observe pulsation.

Equation~\eqref{eq:rho} represents the opening angle of a "fully-filled" open-field line region. Consequently, the width of the radio profile, $w_{r}$, can be calculated from $\rho$, and the geometry depicted by the angles $\alpha$ and $\xi$ via (see \cite{lorimer_handbook_2004})
\begin{equation}
\cos \rho = \cos \alpha \cos \xi + \sin \alpha \sin \xi \cos (w_{r}/2)  .
\end{equation}
This relation serves to compute the observed pulse width~$w_r$ of our sample knowing all other parameters.

\subsubsection{Radio luminosity}

To estimate if our simulated pulsars can be detected in current pulsar surveys, we exactly follow the procedure suggested by \cite{D.Smith} where the
radio flux density $F_r$ of a pulsar at 1.4~GHz is written as,
\begin{equation}\label{eq:Fr}
F_{r} = 9 \, \textrm{mJy} \, \left(\frac{d}{1~\textrm{kpc}}\right)^{-2} \, \left(\frac{\dot{E}}{10^{29}~\textrm{W}}\right)^{1/4} \, 10^{F_{j}}
\end{equation}


where $d$ is the distance in kpc and $F_{j}$ is the scatter term which is modelled as a Gaussian with a mean of $\mu = 0.0$ and a variance of $\sigma = 0.2$. Then, for example, a pulsar with $\dot{E}=10^{29}$~W at a distance of $1~$kpc has a mean flux density of 9~mJy. Thus, if in a radio survey the minimum  flux is $S^{\rm min}_{\rm survey}$, then detection signal-to-noise ratio (S/N) is
\begin{equation}
S/N = \frac{F_r}{S^{\rm min}_{\rm survey}}
\end{equation}
The minimum flux that is related to its period $P$ and width of radio emission $w_r$
\begin{equation}
  S^{\rm min}_{\rm survey} = S_0 \sqrt{\frac{\tilde{w_{r}}}{P-\tilde{w_{r}}}}
\end{equation}
here $\tilde{w_{r}}=w_{r}\,P/2\pi$. The scaling factor $S_0$ reflects the survey parameters. \cite{D.Smith} noted that currently the two most deep pulsar surveys that detected a large number of pulsars are the Parkes multibeam survey (\cite{Parkes2001}) and the Arecibo survey (\citealt{Cordes2006_Arecibo}) where both these surveys for normal pulsars have a sensitivity of $\sim$0.15 mJy. Hence, for detecting a pulsar with S/N$\sim$10, and using $F_r \approx 0.15$ mJy and pulsar width $\tilde{w_r}=0.1 P$, the estimated $S_0 \sim 0.05~$mJy. We use the same criteria as that of \cite{D.Smith} for pulsar detection, and in our simulation, we obtain $F_r$, $P$ and $\tilde{w_r}$ and if the signal-to-noise ratio S/N$> 10$, then we identify the pulsar as detected.

\subsection{Gamma-ray emission model}

Our gamma-ray photon production relies on the emission emanating from the current sheet within the striped wind. The ultra-relativistic plasma velocity directed along the radial direction radiates along the velocity vector within a small cone of half-opening  $1/\Gamma$ where $\Gamma$ is the wind Lorentz factor. The detailed geometry and efficiency of our model is now described.

\subsubsection{Gamma-ray geometry}

As shown by \cite{petri_young_2021}, the striped wind geometry in the split monopole force-free magnetosphere resembles the actual dipole force-free magnetosphere, connecting the current sheet wobbling angle around the equator to the pulsar magnetic obliquity and the gamma-ray peak separation~$\Delta$. The analytical relation reads
\begin{equation}
\label{eq:SeparationPic}
\cos(\pi\,\Delta) = |\cot \xi\, \cot \alpha|.
\end{equation}

A distant observer detects only gamma-ray photons when the current sheet crosses his line of sight. This constrains the line of sight to remain along the equator, with an inclination angle bounded by $|\xi - \pi/2| \leq \alpha$. In our population synthesis, we do not investigate the precise pulse profile, discriminating between single or double peak gamma-ray profile. To complete the gamma-ray emission part, we also need to prescribe the associated luminosity and its possible detection by current instrumentation.

\subsubsection{Gamma-ray luminosity}

\label{eq:Lgamma}

The gamma-ray luminosity function $L_\gamma$ is extracted from the recent study of \cite{Kalapotharakos2019}, where they showed that it is described by a fundamental plane such that $L_{\gamma}=f(\varepsilon_{cut}, B, \dot{E})$
where $\epsilon_{cut}$ is the cut off energy and $B$ the surface magnetic field strength.
Since $\epsilon_{cut}$ is not accessible in our model, we choose their 2D model, not requiring information about the cutoff energy
\begin{equation}\label{Eq:lgamma_2d}
L_{\gamma (2D)} = 10^{26.15 \pm 2.6} \textrm{ W} \, \left( \frac{B}{10^8 \textrm{ T}} \right)^{0.11 \pm 0.05} \, \left(\frac{\dot{E}}{10^{26} \textrm{ W}}\right)^{0.51 \pm 0.09} . 
\end{equation}
This expression remains very similar to the one given by \citep{petri_unified_2011} where he showed that
\begin{equation}
\label{eq:Lgamma}
L_{\gamma} = \dot{E} \sqrt{\frac{10^{26} \textrm{ W}}{\dot{E}}}
\end{equation}
except for the weak dependence on the magnetic field introduced by \cite{Kalapotharakos2019}. We will use expression \eqref{Eq:lgamma_2d} for the gamma-ray luminosity prescription.

The $\gamma$-ray flux as detected on Earth, $F_{\gamma}$, is then simply $L_{\gamma}$ corrected for the line-of-sight cut with beaming fraction~$f_\Omega$ and divided by the square of the distance~$d$ to a given pulsar. Explicitly we get
\begin{equation}
F_{\gamma} = \frac{L_{\gamma (2D)}}{4\,\pi\,f_\Omega\,d^2} . 
\end{equation}
For the striped wind model, the $f_\Omega$ factor has been computed by \cite{petri_unified_2011-1} and varies between $0.22$ and $1.90$. We use a rough approximation for the beaming fraction term~$f_\Omega$ such that if $\alpha < - \xi +0.6109$ then $f_\Omega=1.9$ and $f_\Omega=1$ otherwise. From the sensitivity expectation of the Fermi/LAT instrument, we assume that the sensitivity to pulsars at Galactic latitudes $<2^\circ$ is $F_{\rm min}=\SI{4e-15}{W.m^{-2}}$ and $\SI{16e-15}{W.m^{-2}}$ for blind searches. 


\section{Simulations}
\label{sec:simulations}

In this paper, we have mainly focused our attention on reproducing the form of the spin period versus period derivative $P-\dot{P}$ diagram and the number of pulsar types. To compare our simulations with the observations, we have excluded the binary pulsars and the pulsars with $P < 20~$ms from the ATNF catalogue. There is no need to exclude the pulsars with $P<20~$ms in our simulations, since there are virtually none. Given that the intrinsic parameters provided in section~\ref{sec:model} are poorly constrained, they have been varied within their likely minimum and maximum values until the model satisfactorily reproduces both the $P-\dot{P}$ diagram and the number of radio-only, radio-loud gamma-ray and gamma-ray pulsars (for $\dot{E} >10^{31}$~W and $\dot{E} >10^{28}$~W (see Table.~\ref{tab:simu_Ndetec})). This enables us to constrain the allowed regions in the parameter space of the pulsar's intrinsic parameters. We get a reasonable estimate of the observations with the parameters already given in table \ref{tab:params}. It should be noted that the parameters are changed manually, to save time we do not use a multidimensional root finder algorithm.

  \begin{figure*}
\centering
\begin{subfigure}{.5\textwidth}
  \centering
  \includegraphics[width=1.1\linewidth]{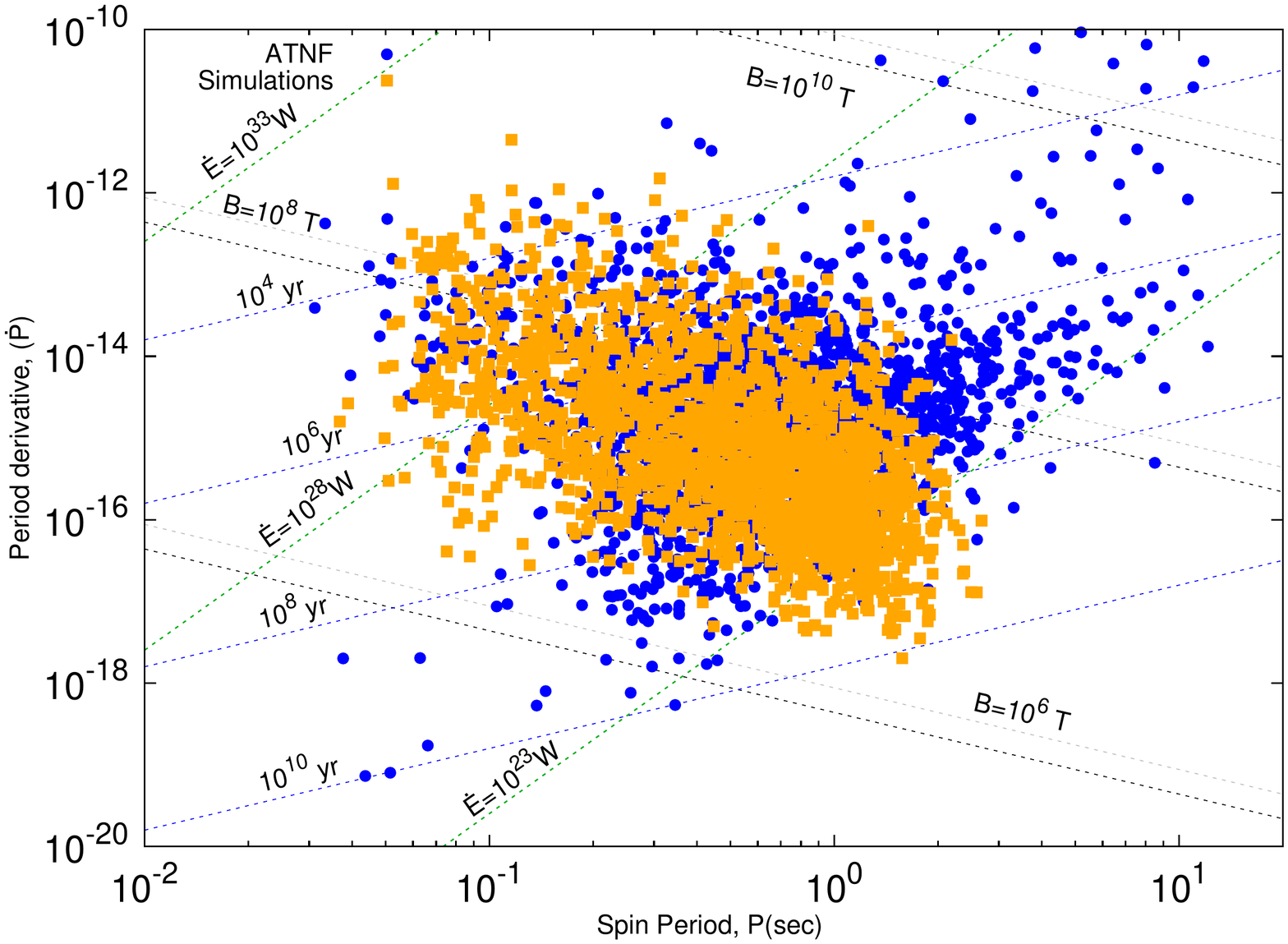}
  \caption{Decaying magnetic field case.}
  \label{fig:sub1}
\end{subfigure}%
\begin{subfigure}{.5\textwidth}
  \centering
  \includegraphics[width=1.1\linewidth]{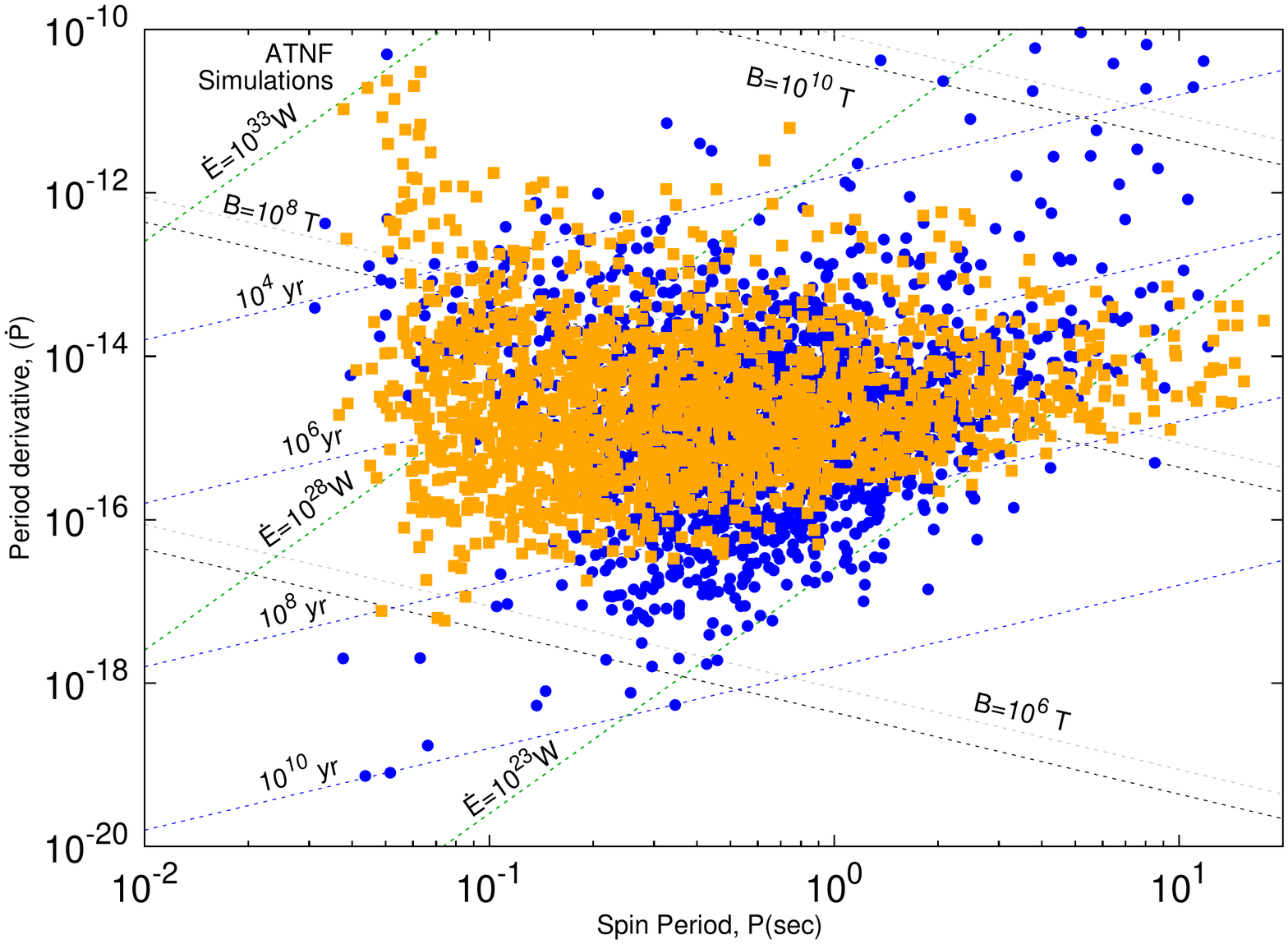}
  \caption{Constant magnetic field case.}
  \label{fig:sub2}
\end{subfigure}
\caption{$P-\dot{P}$ diagram of the simulated population along with the observations including both gamma-ray and radio pulsars.}
\label{fig:ppdot}
\end{figure*}     
              

\begin{figure}
    \centering
    \includegraphics[width=0.5\textwidth]{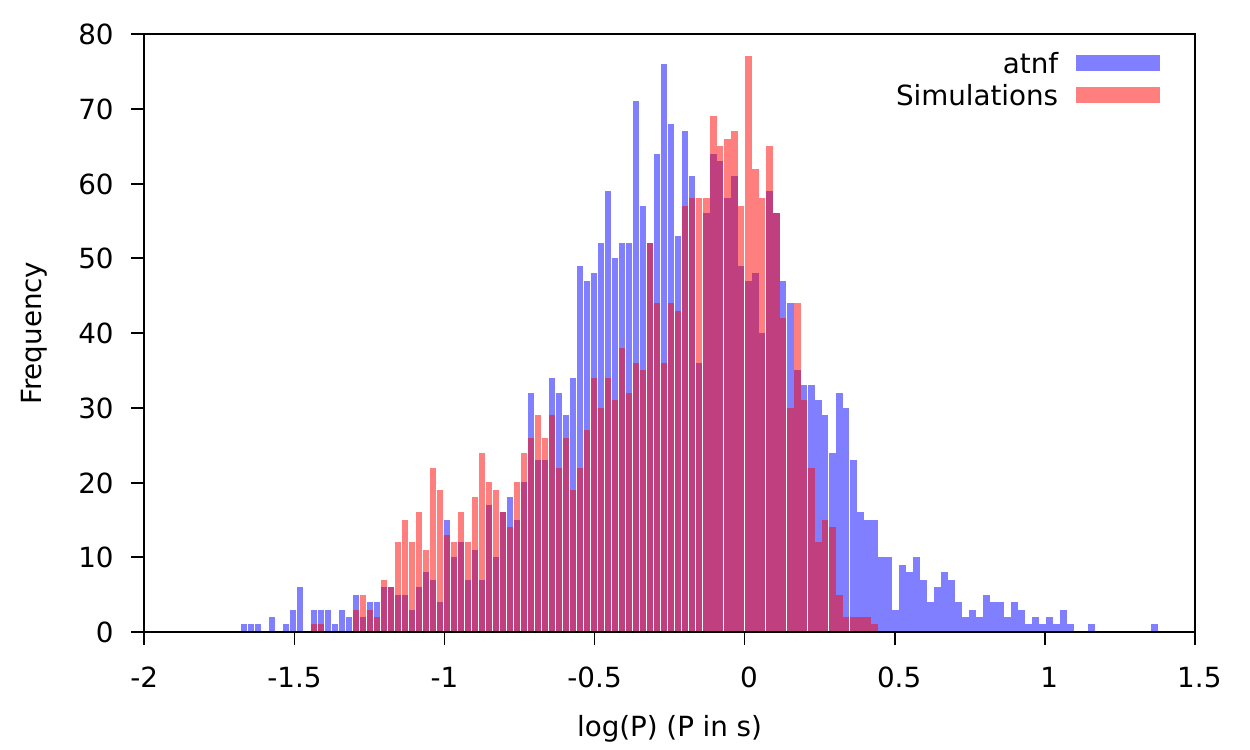}
    \caption{Distribution of the observed period taken from the ATNF catalogue along with the simulations. }
    \label{fig:period}
\end{figure}

\begin{figure}
    \centering
    \includegraphics[width=0.5\textwidth]{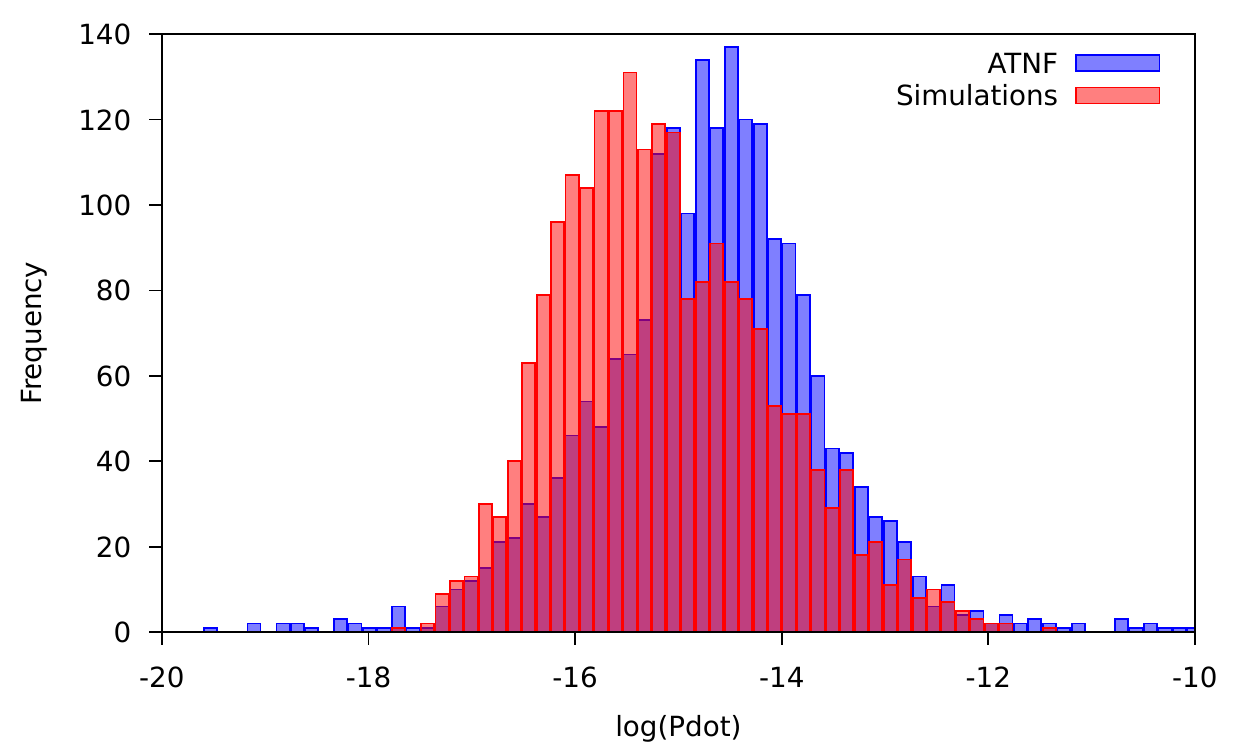}
    \caption{Distribution of the observed period derivative $\dot{P}$ taken from the ATNF catalogue along with the simulations. }
    \label{fig:pdot}
\end{figure}

\subsection{$P-\dot{P}$ diagram}

The resulting $P-\dot{P}$ diagram extracted from our simulations is shown on Fig.~\ref{fig:ppdot} panel (a) for decaying magnetic field and panel (b) for constant magnetic field, along with the pulsar data from ATNF catalog. Those diagrams are obtained with the parameters given in table \ref{tab:params}. As it can be seen, the model, including the magnetic field decay, matches better the observations than the model a constant magnetic field.
We notice that with a decaying magnetic field more pulsars are lying in the lower part of the diagram (meaning smaller $\dot{P}$) whereas for a constant magnetic field more pulsars are lying in the right part of the diagram (meaning longer periods). This is due to the effect of magnetic field evolution, which clearly causes the longer period pulsars to slow down less rapidly than with a constant magnetic field. This mismatch could be improved by refining the distributions of periods and magnetic fields at birth, but other parameters could also impact the current $P-\dot{P}$ distribution like for instance the detection limits. The radio emission mechanism can also affect this discrepancy. The exact nature of these dependencies that accurately represents the observed pulsar population needs a detailed investigation, which is beyond the scope of this work. In the $P-\dot{P}$ diagram, the region with a magnetic field greater than $\sim 10^8~$T and $10^{23} W <\dot{E} < 10^{28} W$ is also depleted. Indeed, the high magnetic field pulsars are not taken into account in our study because they probably belong to a distinct class of magnetized neutron stars. For a better representation of the $P-\dot{P}$ diagram, we also show the histograms of periods $P$ in Fig.~\ref{fig:period} and period derivatives $\dot{P}$ in Fig.~\ref{fig:pdot} along with the data.

         \begin{figure}
         \includegraphics[width=9.5cm,left]{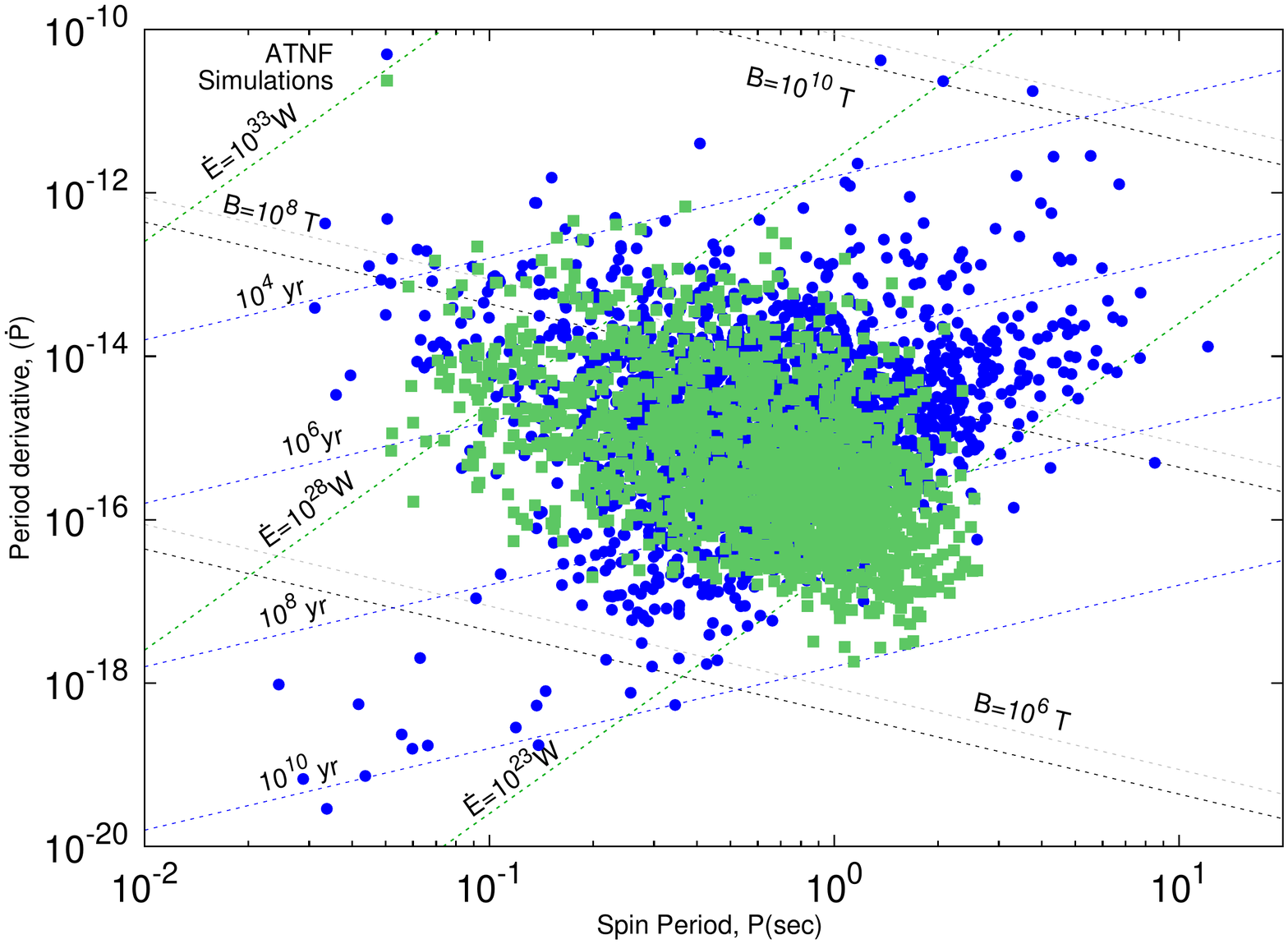}
          \caption{$P-\dot{P}$ diagram of the radio-only pulsars for both the simulations and the observations. }
   \label{Fig:radio}
       \end{figure}
       
       \begin{figure}
         \includegraphics[width=9.5cm,left]{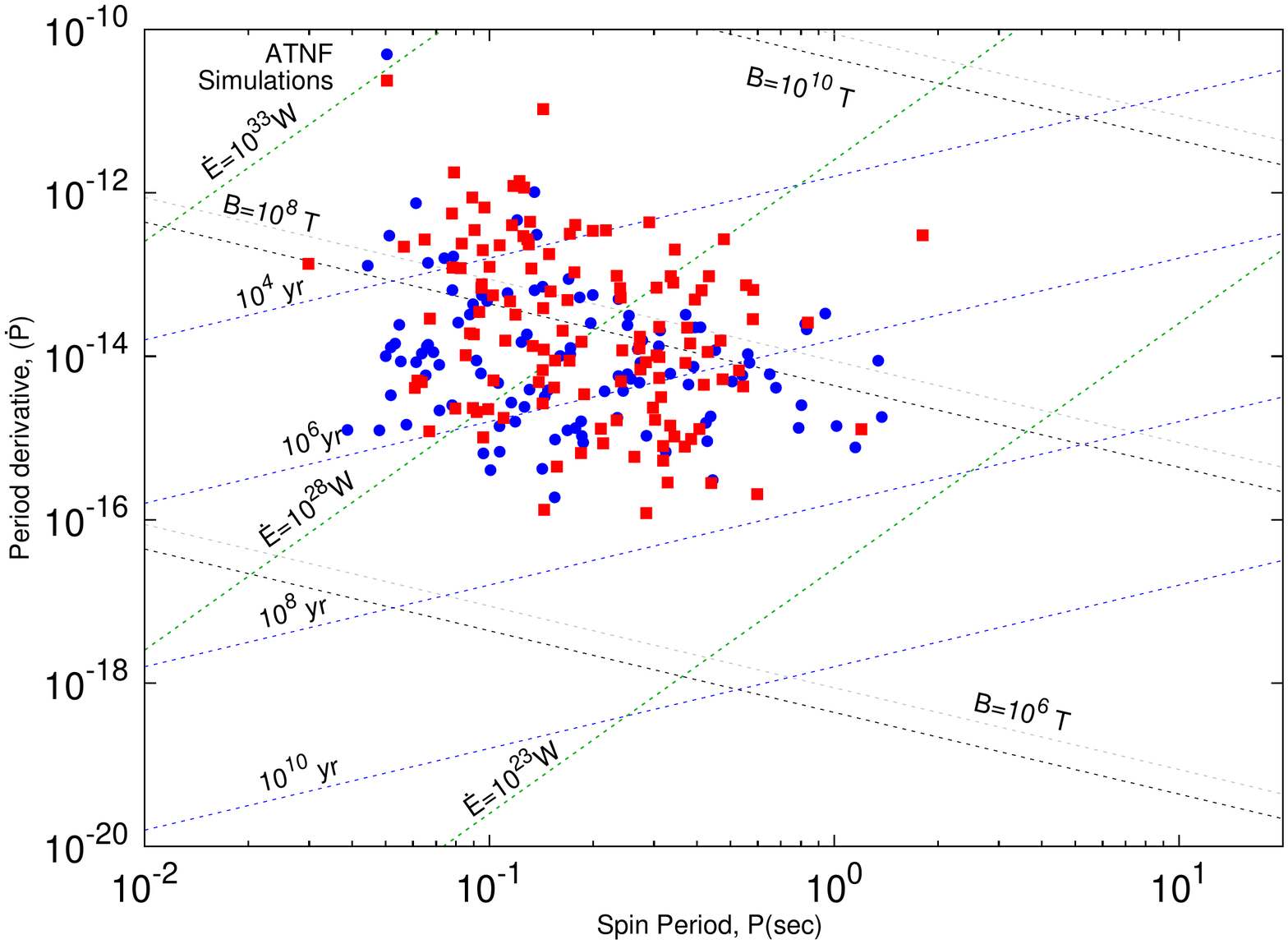}
          \caption{$P-\dot{P}$ diagram of the gamma-only pulsars for both the simulations and the observations. }
   \label{Fig:gamma}
       \end{figure}
  \begin{figure}
         \includegraphics[width=9.5cm,left]{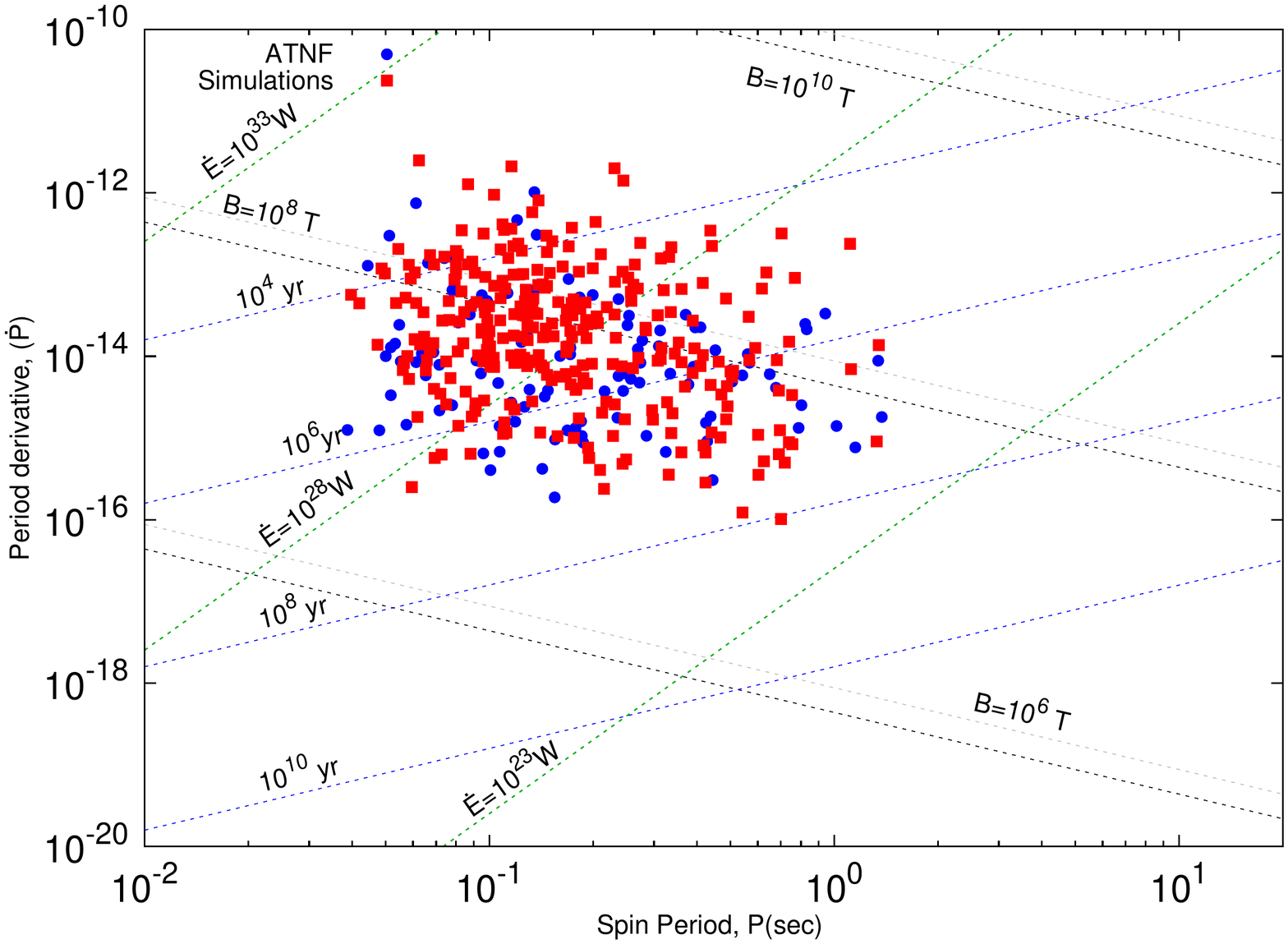}
          \caption{$P-\dot{P}$ diagram of the radio-loud gamma ray pulsars for both the simulations and the observations.}
   \label{Fig:radio_gamma}
       \end{figure}
      

\subsection{Pulsar types}

We have computed the total number $N_{r}$, $N_{g}$, $N_{rg}$ that are the number of radio-only, gamma-only and radio-loud gamma-ray pulsars respectively, and within the energy bands $\dot{E}=10^{28}$~W and $\dot{E}=10^{31}$~W. The resulting numbers are within the observations, and are better constrained by the decaying magnetic field model (see Table~\ref{tab:simu_Ndetec_bconst}) and Table~\ref{tab:simu_Ndetec}. 
We have represented the $P-\dot{P}$ diagram for $N_r$, $N_g$ and $N_{rg}$ on Fig.~\ref{Fig:radio}, Fig.~\ref{Fig:gamma}, and Fig.~\ref{Fig:radio_gamma} for both the simulations and the observations. We find good agreement between the observed and simulated population separately for each subclass.

\begin {table}[H] 
\small\tabcolsep=0.15cm
\begin{tabular}{l l l l l}
 \hline
 \hline
  $\log(\dot{E})$ (in W) &$N_{\rm tot}$ & $N_{\rm r}$ & $N_{\rm g}$ & $N_{\rm rg}$ \\  
       \hline
$> 31 $ & 25 & 1  & 15  & 9   \\
$> 28 $ & 298 & 100  & 73  & 125  \\
 \hline 
 total & 1882 & 1553 & 136 & 193 \\
\hline
\end{tabular}
\caption{The quantities $N_{r}$, $N_{g}$, $N_{rg}$ are the number of radio-only, gamma-only and radio-loud gamma-ray pulsars respectively obtained from our simulations for a constant magnetic field.} 
\label{tab:simu_Ndetec_bconst}
\end {table}

\begin {table}[H] 
\small\tabcolsep=0.15cm
\begin{tabular}{l l l l l}
 \hline
 \hline
  $\log(\dot{E})$ (in W) &$N_{\rm tot}$ & $N_{\rm r}$ & $N_{\rm g}$ & $N_{\rm rg}$ \\  
       \hline
$> 31 $ & 3 & 0  & 1  & 2   \\
$> 28 $ & 238 & 87  & 47  & 104  \\
 \hline 
 total & 2155 & 1864 & 122 & 169 \\
\hline
\end{tabular}
\caption{The quantities $N_{r}$, $N_{g}$, $N_{rg}$ are the number of radio-only, gamma-only and radio-loud gamma-ray pulsars respectively obtained from our simulations with a decaying magnetic field.} 
\label{tab:simu_Ndetec}
\end {table}


\subsection{Spatial distribution}

\begin{figure}
    \centering
     \includegraphics[width=0.5\textwidth]{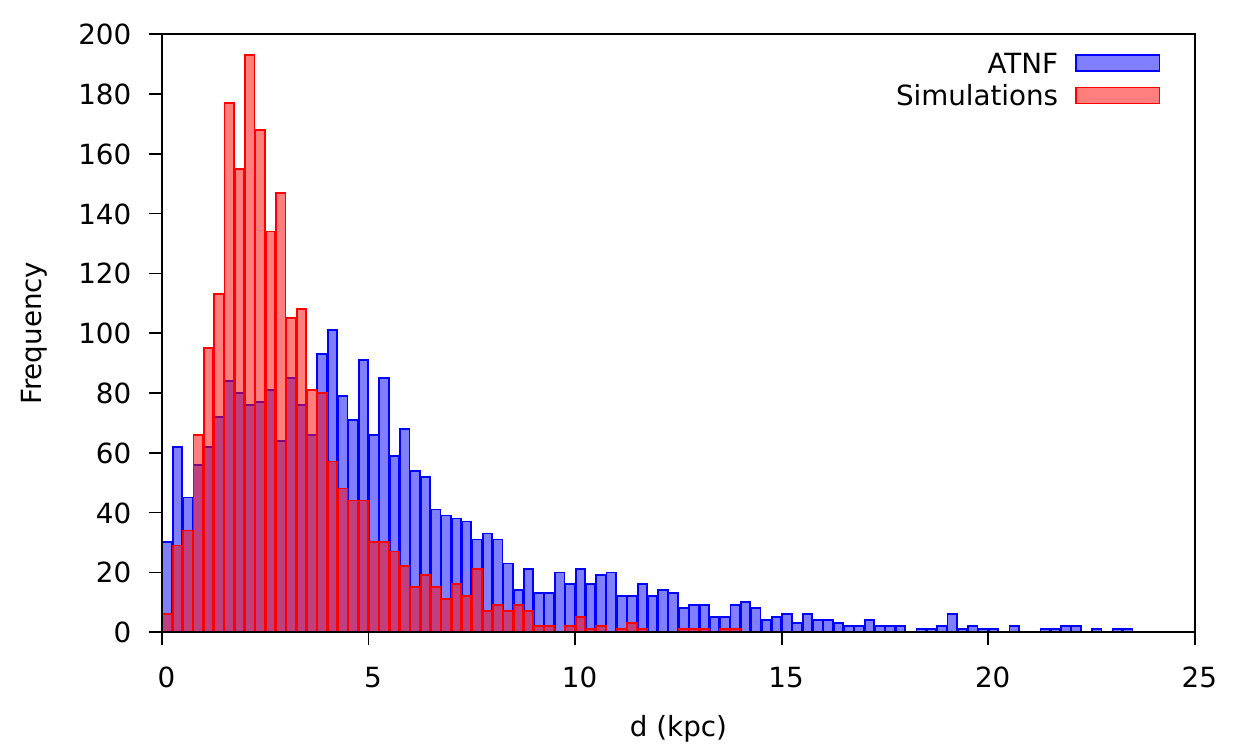}
    \caption{ Distribution of the distance to Earth for the simulated and observed populations. }
    \label{fig:distance}
\end{figure}

\begin{figure}
    \centering
    \includegraphics[width=0.5\textwidth]{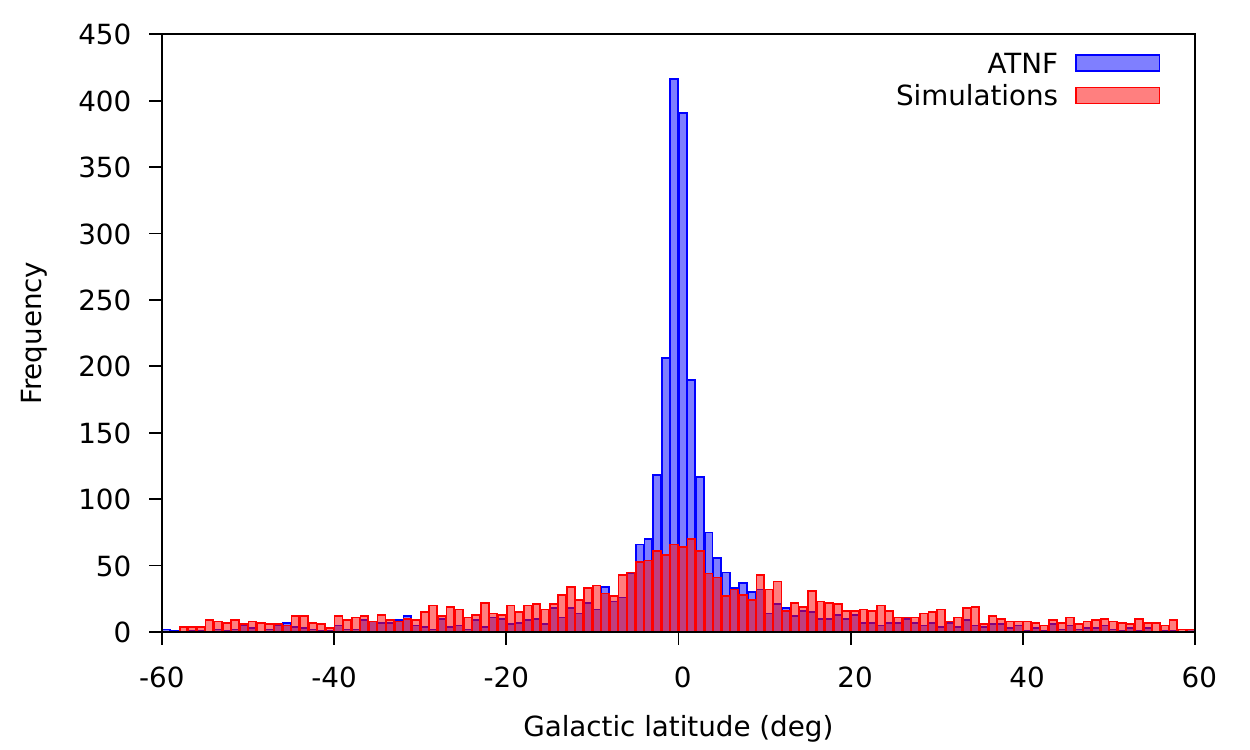}
    \caption{Distribution of the observed galactic latitude, $b$ in degrees taken from the ATNF catalogue along with the simulations. }
    \label{fig:latitude}
\end{figure}

\begin{figure}
 \includegraphics[width=0.5\textwidth]{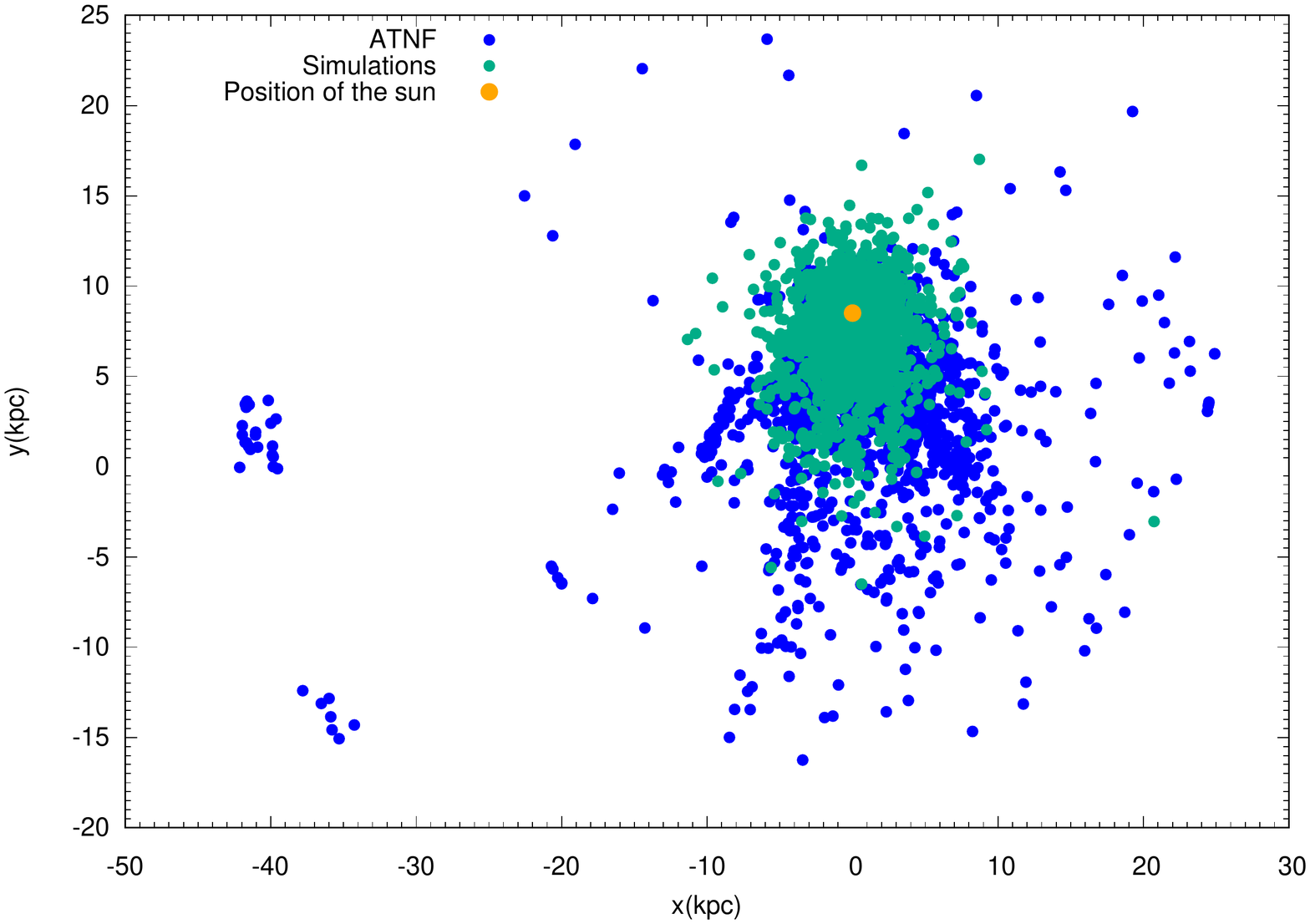}
  \caption{Detected and observed spatial distribution of pulsars projected onto the galactic plane. }
  \label{Fig:xy}
\end{figure}

The distance $d$ from Earth to the detected pulsars is shown in Fig.~\ref{fig:distance} for both the simulations and the observations. We found that more pulsars are detected at closer distances, and less at higher distances. The projected 2D distribution is shown in Fig.~\ref{Fig:xy} which represents the galactic coordinate $y$ as a function of $x$. As it was already mentioned earlier, since we detect more pulsars with low $\dot{E}$, there we detect more pulsars that are closer to us (see Eq.~\ref{eq:Fr}). The distribution of the latitude for the simulated and observed population is shown on Fig.~\ref{fig:latitude}. 
In our simulations, we detect fewer pulsars at lower galactic latitude. This is due to the spatial distribution of pulsars, which shows pulsars closer to the Sun compared to observations and therefore artificially increasing their latitude. This may indicate that a more refined treatment of the sensibility depending on the latitude, and a better description of the initial spatial distribution as well as the birth kick velocity should be implemented. Furthermore, our current simulation does not take into account the galactic potential. The simulated pulsar's latitude spread would shift more towards the galactic plane, thus increasing the number of pulsars in lower galactic latitudes, and perhaps better representing the data.

\subsection{Influence of the parameters}

Here we want to show the influence of the parameters on the quantities $N_r$, $N_g$ and $N_{rg}$ that are the number of radio only, gamma only and radio-loud gamma-ray pulsars respectively from our simulations with the parameters given in table~\ref{tab:params}. This is summarized on tables \ref{tab:brate}, \ref{tab:B}, \ref{tab:period}, \ref{tab:sigmap}, \ref{tab:sigb} and \ref{tab:S_N}.
The most sensitive parameters are the birth rate, the initial mean magnetic field and $\sigma_b$. 

First the number of detected pulsars~$N_{\rm tot}$ scales linearly with the birth rate~$\tau_{\rm birth}$. Increasing the birth rate by a factor 2 will decrease~$N_{\rm tot}$ by the same factor. We indeed detect statistically more pulsars if more are present within our galaxy. This trend is clearly seen for the rates 1/50yr, 1/100yr and 1/200yr, see table~\ref{tab:brate}.
\begin {table} 
\begin{flushleft}
\small\tabcolsep=0.15cm
\begin{tabular}{l l l l l}
 \hline
 \hline
  birth rate  (in $yr^{-1}$) &$N_{\rm tot}$ & $N_{\rm r}$ & $N_{\rm g}$ & $N_{\rm rg}$ \\  
       \hline
$ 50 $ & 3080 & 2635  & 191  & 254   \\
$ 100 $ & 1517 & 1306  & 87  & 124  \\
$ 200 $ & 752 & 655  & 41  & 56  \\
$ 300 $ & 489 & 423  & 34  & 32  \\

 \hline 

\end{tabular}
\end{flushleft}
\caption{Influence of the birth rate on the quantities $N_{r}$, $N_{g}$, $N_{rg}$. 
\label{tab:brate}}
\end {table}  


The mean magnetic field also drastically impacts~$N_{\rm tot}$. The number of detected pulsars decreases with increasing field strength. This is mainly due to the fact that high magnetic pulsars align faster than low magnetic field pulsars. Furthermore, the magnetic field decay is also faster for higher magnetic fields.

\begin {table} 
\begin{flushleft}
\small\tabcolsep=0.15cm
\begin{tabular}{l l l l l}
 \hline
 \hline
  $B_{mean}$  (in T) &$N_{\rm tot}$ & $N_{\rm r}$ & $N_{\rm g}$ & $N_{\rm rg}$ \\  
       \hline
$ 10^7 $ & 16713 & 15051  & 398  & 1264   \\
$ 2.5\times10^7 $ & 10597 & 9306  & 369  & 922  \\
$ 5\times10^7 $ & 6887 & 5952  & 278  & 657  \\
$ 10^8 $ & 4272 & 3683 & 212  & 377  \\
$ 5\times 10^8 $ & 1298 & 1131  & 67  & 100  \\
\hline
\end{tabular}
\end{flushleft}
\caption{Influence of the initial magnetic field on the quantities $N_{r}$, $N_{g}$, $N_{rg}$ that are the number of radio only, gamma only and radio-loud gamma-ray pulsars respectively from our simulations with the parameters given in table~\ref{tab:params}.
\label{tab:B}}
\end {table}  

The initial mean period moderately impacts the detected pulsar number, as shown in table~\ref{tab:period}. Increasing this period by a factor~10 only increases $N_{\rm tot}$ by a factor~2.
 \begin {table} 
\begin{flushleft}
\small\tabcolsep=0.15cm
\begin{tabular}{l l l l l}
 \hline
 \hline
  $P_{mean}$  (in s) &$N_{\rm tot}$ & $N_{\rm r}$ & $N_{\rm g}$ & $N_{\rm rg}$ \\  
       \hline
$ 10 $ & 1145 & 856  & 156  & 133   \\
$ 20 $ & 1509 & 1195  & 134  & 180  \\
$ 50 $ & 1914 & 1638  & 116  & 160  \\
$ 100 $ & 2491 & 2235  & 104  & 152  \\
\hline
\end{tabular}
\end{flushleft}
\caption{ Influence of the initial period on the quantities $N_{r}$, $N_{g}$, $N_{rg}$. 
\label{tab:period}}
\end {table}  

Moreover, $N_{\rm tot}$ remains insensitive to the spread in this birth period, see table~\ref{tab:sigmap}.
 \begin {table} 
\begin{flushleft}
\small\tabcolsep=0.15cm
\begin{tabular}{l l l l l}
 \hline
 \hline
  $\sigma_{P}$  (in s) &$N_{\rm tot}$ & $N_{\rm r}$ & $N_{\rm g}$ & $N_{\rm rg}$ \\  
       \hline
$ 0.001 $ & 2071 & 1787  & 133  & 151   \\
$ 0.005 $ & 2178 & 1918  & 111  & 149  \\
$ 0.01 $ & 2119 & 1827  & 125  & 167  \\
$ 0.05 $ & 2204 & 1916  & 133  & 155  \\
\hline
\end{tabular}
\end{flushleft}
\caption{Influence of the $\sigma_{P}$ on the $N_{r}$, $N_{g}$, $N_{rg}$. 
\label{tab:sigmap}}
\end {table}  

Increasing the spread in magnetic field strength allows for lower B field values and, as seen above, a larger number of pulsars will be detected. 
In this paper, we have used $S/N=10$ to get the detected pulsars, however, in reality pulsars have been detected with telescopes having different sensitivities. The table \ref{tab:S_N} allows to understand better how the detection changes with the signal-to-noise ratio.

Decreasing the $S/N$ ratio naturally increases the number of detected pulsars. For instance, switching from a $S/N=10$ to a $S/N=5$ multiplies the number of detected pulsars by a factor $1.7$.
 \begin {table}[H] 
\begin{flushleft}
\small\tabcolsep=0.15cm
\begin{tabular}{l l l l l}
 \hline
 \hline
 $\sigma_b$  (in T) &$N_{\rm tot}$ & $N_{\rm r}$ & $N_{\rm g}$ & $N_{\rm rg}$ \\  
       \hline
$ 0.1 $ & 1500 & 1290  & 122  & 88   \\
$ 0.2 $ & 1616 & 1369 & 138  & 109\\
$ 0.3 $ & 1661 & 1467  & 90  & 104  \\
$ 0.4 $ & 1923 & 1645  & 111  & 167  \\
$ 1 $ & 4427 & 3910  & 168  & 349  \\
\hline
\end{tabular}
\end{flushleft}
\caption{Influence of $\sigma_b$ on the quantities $N_{r}$, $N_{g}$, $N_{rg}$. 
\label{tab:sigb}}
\end {table}

\begin {table}[H] 
\begin{flushleft}
\small\tabcolsep=0.15cm
\begin{tabular}{l l l l l}
 \hline
 \hline
 $S/N$ &$N_{\rm tot}$ & $N_{\rm r}$ & $N_{\rm g}$ & $N_{\rm rg}$ \\  
       \hline
$ 3 $ & 8312 & 7960  & 124  & 228   \\
$ 5 $ & 4595 & 4267  & 137  & 191  \\
$ 10 $ & 2155 & 1864  & 122  & 169  \\
$ 15 $ & 2097 & 1911  & 109  & 77  \\

 \hline
\end{tabular}
\end{flushleft}
\caption{Influence of the signal-to-noise ratio on the quantities $N_{r}$, $N_{g}$, $N_{rg}$.} 
\label{tab:S_N}
\end {table}


\subsection{Radio width $w_r$ as a function of the period}

\begin{figure}
 \includegraphics[width=0.5\textwidth]{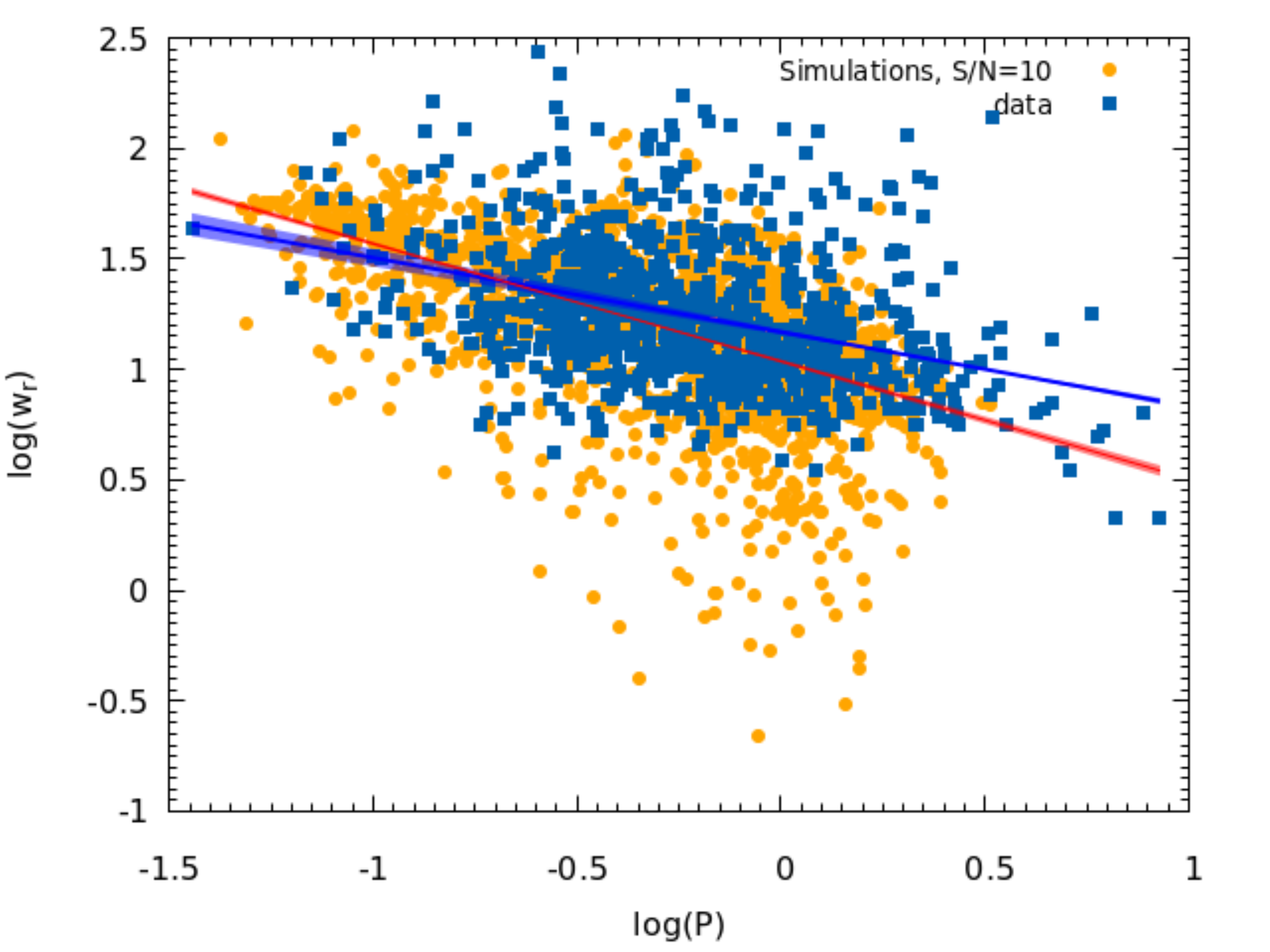}
  \caption{ $\log (w_r)$ as a function of $\log(P)$. The quantity $w_r$ is expressed in degrees and $P$ is expressed in $s$. The data are the width at $10\%~$level taken from \cite{Posselt_2021}. The blue solid line and the shaded area represent the result of a linear fit to the data and its uncertainties and the red solid line and the shaded area represents a linear fit to the simulations with its uncertainties.}
  \label{Fig:wr_P}
\end{figure}

Fig.~\ref{Fig:wr_P} represents the logarithm of the radio pulse profile width $\log(w_r)$ as a function of the logarithm of the period $\log(P)$, to compare our simulations with the data. We retrieve the trend shown in the observations, namely a decrease of the width with an increase in the period. The slope in log-log scale is $-0.53 \pm 0.02$, meaning $w_r = (1.04 \pm 0.01)P^{-0.53 \pm 0.02}$ as expected by the emission cone model. A linear fit to the measured pulsar width data from \cite{Posselt_2021} gives $w_{10}=(1.17 \pm 0.01)P^{-0.34 \pm 0.03}$, which, although indicative of the expected trend, still does not agree with the index of $\sim -0.5$. We believe that a more robust analysis and a larger data set will be more appropriate to verify the width period relation, which is beyond the scope of this work. 
However, though expected from geometrical considerations, no small values for $w_r$ have been obtained, particularly for faster periods. This may be explainable by the fact that fast rotating pulsars usually show larger opening angle for the emission beam, naturally leading to a broadening of the pulse profile.

\subsection{Inclination angle}

\begin{figure}
    \centering
    \includegraphics[width=0.5\textwidth]{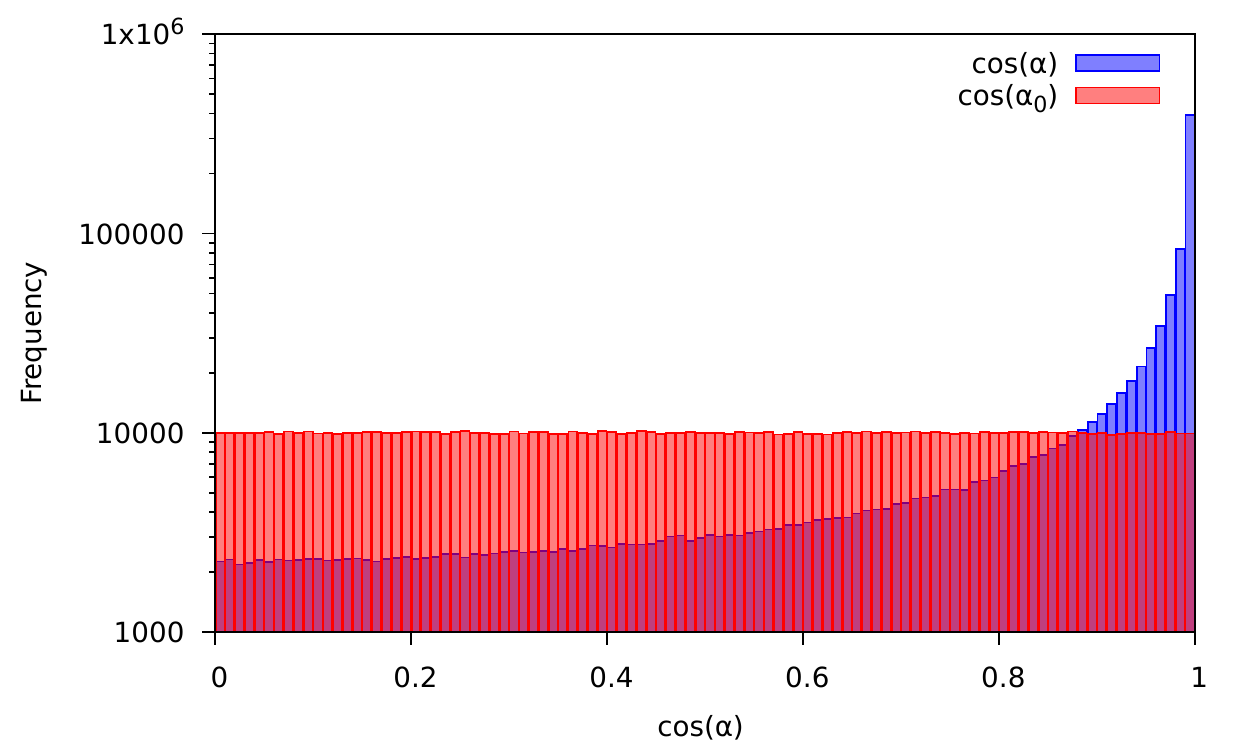}
    \caption{Distribution of $\cos \alpha$ and $\cos \alpha_0$ for a subsample of $10^6$ simulated population (\textit{before} the detection).}
    \label{Fig:alpha_init}
\end{figure}

\begin{figure}
    \centering
    \includegraphics[width=0.5\textwidth]{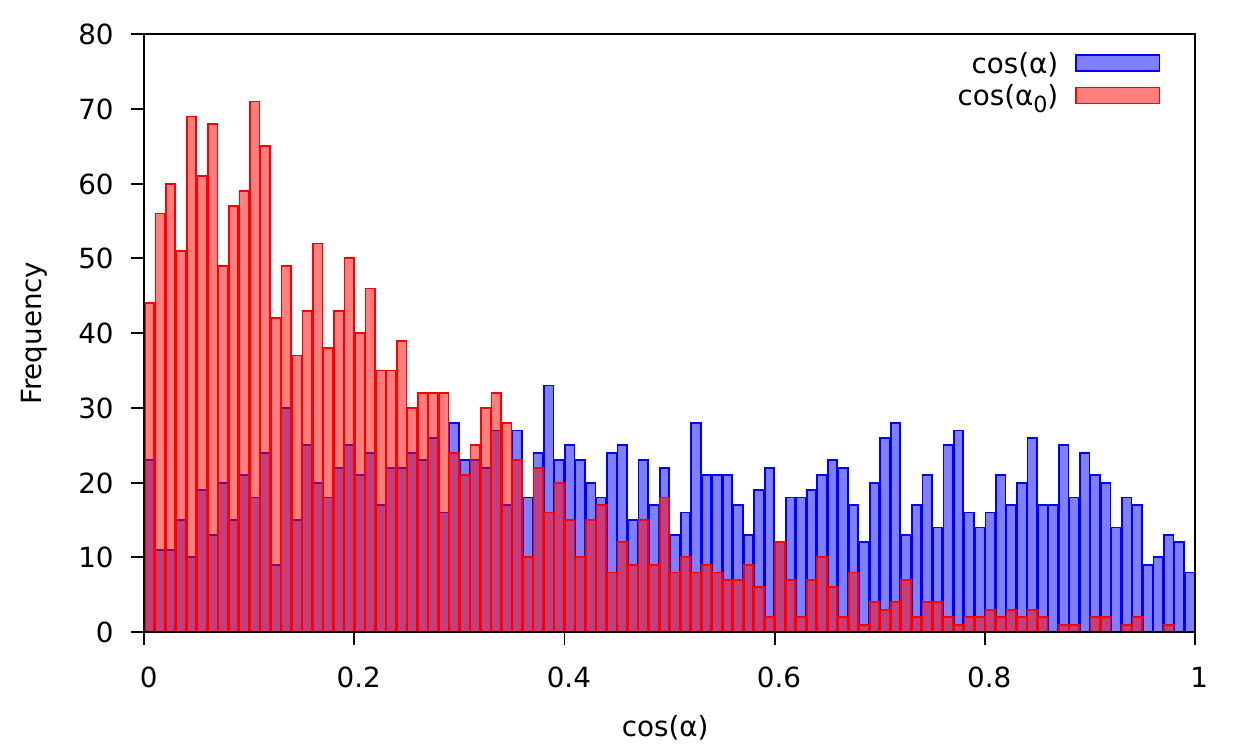}
    \caption{ Distribution of $\cos(\alpha)$ and $\cos \alpha_0$ for the \textit{detected} pulsars. }
    \label{Fig:alpha_detec}
\end{figure}

\begin{figure}
    \centering
    \includegraphics[width=0.5\textwidth]{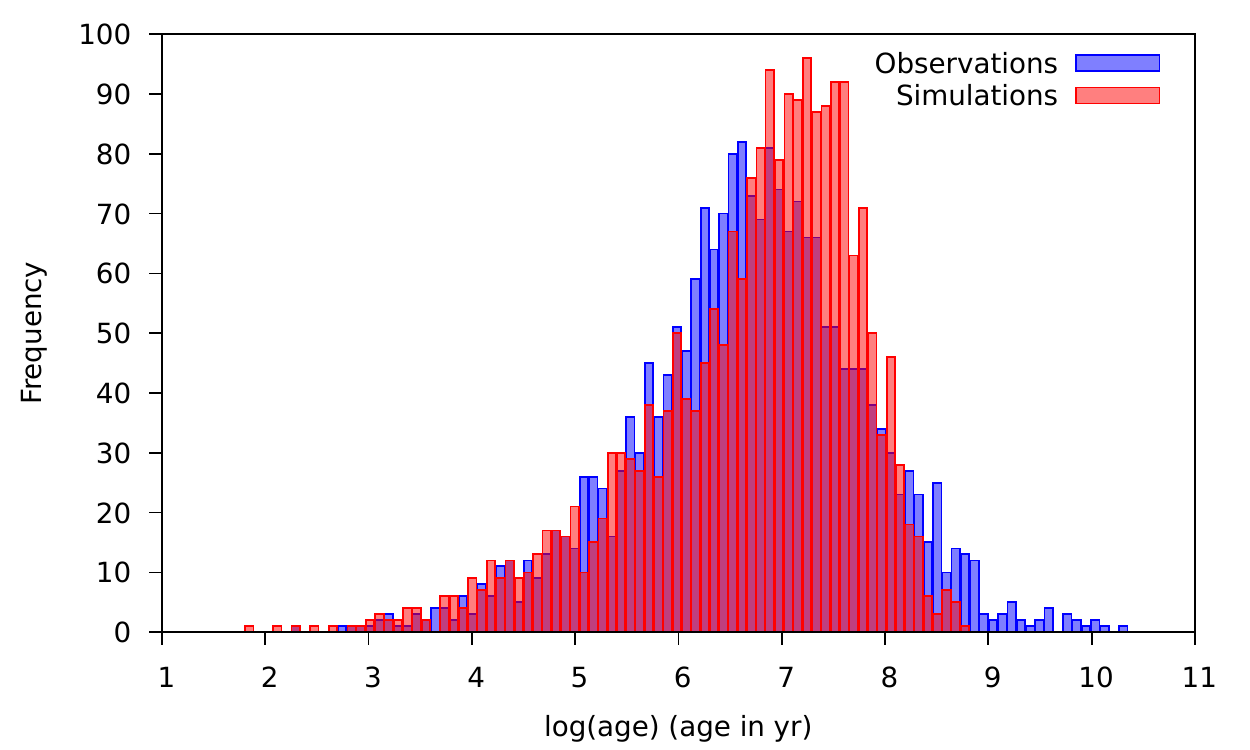}
    \caption{Age distribution of the detected and observed pulsars. }
    \label{Fig:age_distrib}
\end{figure}

\begin{figure}
    \centering
    \includegraphics[width=0.5\textwidth]{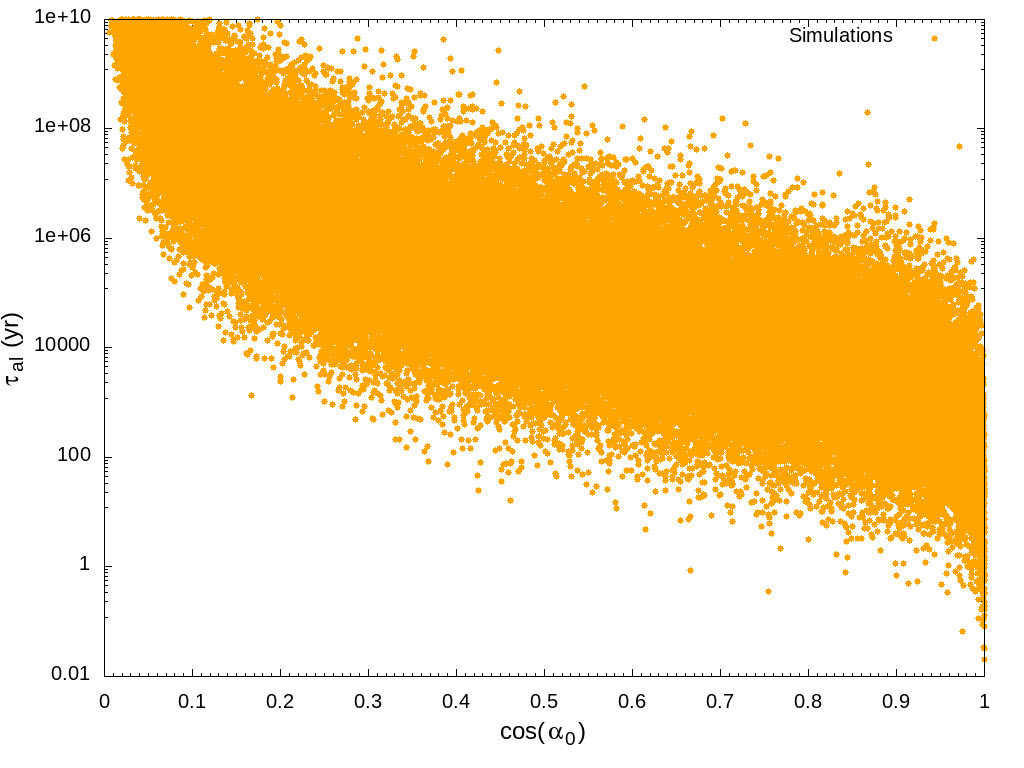}
    \caption{Alignment timescale, $\tau^{MHD}_{align}$ as a function of $\cos(\alpha_0)$. Simulation for $10^7$ pulsars. }
    \label{Fig:tau}
\end{figure}

Figure~\ref{Fig:alpha_init} shows the initial obliquity $\cos(\alpha_0)$ at birth and  the current obliquity $\cos (\alpha)$ distribution of the pulsars, irrespective of their detection. Most of the pulsars tend to the aligned configuration when aging. This is due to the fact that older pulsars are in a larger number and are more likely to be aligned. Indeed, Fig.~\ref{Fig:age_distrib} highlights that most of the detected pulsars in our sample are older than $10^6$~yr. This is expected since the birth rate is constant up to Gyr. The Figure~\ref{Fig:alpha_detec} shows the distributions of initial $\cos(\alpha_0)$ and current $\cos (\alpha)$ obliquity for the detected pulsars. The current obliquity distribution $\cos \alpha$ is almost uniform, whereas the distribution at birth $\cos(\alpha_0)$ peaks for values less than $0.4$. This means that mostly only pulsars with larger initial inclination angles, close to orthogonal rotators, will be detected. Moreover, the alignment timescale is longer for high obliquities meaning small $\cos \alpha_0$ (see Fig.~\ref{Fig:tau}). Since when the pulsar approaches an aligned rotator, the pulsed emission becomes harder to detect, pulsars with a longer alignment timescale are favored, hence a bias towards small $\cos \alpha_0$.

\section{Discussion}
\label{sec:discussion}


Population synthesis of neutron stars represents a valuable and reliable tool to constrain the basic physics of neutron star electrodynamics, from its surface up to large distances well outside the light-cylinder. Following the work of \cite{Gullon2014}, we have shown that a realistic pulsar model such as the force-free evolution scenario can self-consistently account for the observed population of canonical pulsars and hence verified the findings of \cite{Gullon2014}. 
Additionally, we have considered the subpopulation of gamma-ray pulsars associated with the striped wind emission model. We have found a birth rate of 1/70~yr which is consistent with previous studies \citep{Gonthier_beam_pulse_profiles2004}, \citep{WattersRomani}. In the case of a constant magnetic field, we found that a birth rate of 1/300~yr improved the matching to the observations. However, this value is not consistent with expectations from previous works. The best results are obtained with a decaying magnetic field model.

 
In our study, we focused only on canonical pulsars for several reasons. First, the radio emission height and geometry are well constrained only for pulsars with periods $P > 20$~ms. Second, the binary pulsars are mostly revived through an accretion phase that is not included in our model. Third, magnetars with super-critical magnetic field strengths require a special initial magnetic field distribution deviating from the normal canonical pulsar population we used. These magnetars are probably produced by a dynamo effect right at their birth during the proto-neutron star stage, increasing their internal magnetic field by several orders of magnitude through turbulent motion or magneto-rotational instabilities. Although doable, we did not include such processes that would lead to a second distribution of magnetic field strengths designed especially for magnetars.

The spatial distribution of neutron stars at birth was postulated and adapted from the literature to fit as well as possible the current observed galactic distribution, taking into account selection effects related to the sensitivity of radio and high energy telescopes. We found a reasonable agreement between the current spatial distribution and the simulated distribution. 


The radio emission geometry constrained by the dipolar region combined with the striped wind emission model for gamma-rays are able to reproduce the entire population of young pulsars summarized in the $P-\dot{P}$ diagram. The number of radio only, gamma-ray only or radio-loud gamma-ray pulsars fits the observed population as given by the ATNF catalog. In a previous study, \cite{PetriMitra2021} demonstrated that the radio-loud gamma-ray pulsar emission model accurately constrained the geometry of the magnetic axis and line of sight axis with respect to the rotation axis. In this paper, we went one step further, showing that our model is consistent with the entire canonical pulsar population.


There are few other selection effects that can affect the detection of a neutron star as a radio pulsar, which we have not included in our analysis. One such selection effect is the propagation of the radio signal in the interstellar medium which is subject to scattering effect that smears out the signal, rendering the detection of some pulsars very difficult because the pulsation disappears at low frequencies. Interstellar scattering is associated with the fluctuation of the electron distribution along the observer's line of sight, where the pulsar signal suffers a multi-path scattering as the signal traverses the ISM, see \cite{Scheuer1968}. The scattering effect, hence can be described as the pulse width to be convolved with the response function of the ISM and as a result the observed pulse width appears to be smeared. If the smearing exceeds the pulse period, then the pulsar cannot be detected as a pulsed signal. The effect of scattering is known to increase with pulsar dispersion measure (see, e.g. \cite{Krishnakumar2015}) and hence, also the distance, thus farther the pulsars it becomes more difficult to detect a pulsar. The scattering timescale is a strong function of observing frequency~$\nu$ with the timescale decreasing as $\sim \nu^{-4.4}$ (see, e.g. \citep{loehmer2004}), and hence sensitive pulsar searches at higher frequencies can in principle be used to detect more pulsars. However, due to steep radio pulsar spectra, pulsars are weaker at higher frequencies and hence it may not be viable to detect a highly scattered pulsar.

The other selection effect is related to the coherent radio emission mechanism for radio pulsars. Our analysis here considers the geometrical model of a star centered magnetic dipole for the radio pulsars, however the condition under which the radio emission mechanism can operate in pulsars are not included. Various studies suggest that the presence of surface multipolar field facilitates the generation of pair plasma, which in turn can assist mechanisms of coherent radio emission. Thus, the nature of the multipolar surface field can affect the number of detectable pulsars, and several studies suggest a death-line in the $P-\dot{P}$ diagram (see e.g. \cite{ChenRuderman1993}; \cite{Mitra2020}). In the current study, however, the pulsars detected via simulations are similar to the actual pulsars, and hence the above selection effects do not appear to have any major impact on the number of detected pulsars. 



\section{Summary}
\label{sec:summary}

In this paper, we showed that the secular evolution of the neutron star spin rate and its braking is accurately captured by the force-free magnetosphere model, in which radio photons emanate from regions near the polar cap whereas the gamma-ray photons are produced within the current sheet of the striped wind. Following simple prescriptions for the birth kick, the birth period and birth magnetic field strength, assuming an isotropic obliquity and an isotropic viewing angle, we were able to pin down the essential parameters of the canonical pulsar population. Almost orthogonal rotators at birth represent the vast majority of pulsars currently detected due to the alignment effect, the geometry of the radio and gamma-ray beams.

Our investigation could be improved in several ways, as exposed in the previous section~\ref{sec:discussion}. Moreover, knowing the geometry of each individual pulsar, we can model the thermal X-ray emission from the polar caps like for instance for PSR~J1136+1551 \citep{petri_joint_2020}. The upcoming advent of the SKA telescope will increase by one order of magnitude the number of known pulsars and constrain even more the evolution scenario and emission physics of these stellar remnants.
\begin{acknowledgements}
This work has been supported by the CEFIPRA grant IFC/F5904-B/2018 and ANR-20-CE31-0010. We acknowledge the High Performance Computing Centre of the University of Strasbourg for supporting this work by providing scientific support and access to computing resources. We thank David Smith for stimulating discussions. DM acknowledges the support of the Department of Atomic Energy, Government of India, under project no. 12-R\&D-TFR-5.02-0700.
\end{acknowledgements}

\newpage

%
%

  \bibliographystyle{aa} 
   
 \bibliography{biblio}

\end{document}